\documentclass[pra, amsmath,amssymb,aps,10pt,twocolumn]{revtex4-1}
\usepackage{graphicx}
\usepackage[usenames, dvipsnames]{color}
\usepackage{dcolumn}
\usepackage{bm}
\usepackage{color}
\usepackage[colorlinks,bookmarks=false,citecolor=blue,linkcolor=blue,urlcolor=black]{hyperref}

\begin{document}

\preprint{APS/123-QED}

\title{Restricted space {\it ab initio} models for double ionization by strong laser pulses}
\author{Dmitry K. Efimov$^1$,
	Artur Maksymov$^1$, Jakub S. Prauzner-Bechcicki$^1$, Jan H. Thiede$^2$, Bruno Eckhardt$^2$, Alexis Chac\'on$^{3,4}$, Maciej Lewenstein$^{4,5}$, and Jakub Zakrzewski$^{1,6}$}
\email{dmitry.efimov@uj.edu.pl}
\affiliation{\mbox{$^1$Instytut Fizyki imienia Mariana Smoluchowskiego, Uniwersytet Jagiello\'nski, \L{}ojasiewicza 11, 30-348 Krak\'ow, Poland}
	\mbox{$^2$Philipps-University Marburg, Biegenstra{\ss}e 10, 35037 Marburg, Germany}\hfill
	\mbox{$^3$Center for Nonlinear Studies, Theoretical Division, Los Alamos National Laboratory, Los Alamos NM 87545, USA}\hfill
	\mbox{$^4$ICFO-Institut de Ciencies Fotoniques, The Barcelona Institute
		of Science and Technology, 08860 Castelldefels (Barcelona), Spain}
	\mbox{$^5$ICREA, Passeig de Lluis Companys, 23, 08010 Barcelona, Spain}
	\mbox{$^6$Mark Kac Complex Systems Research Center, Jagiellonian University, \L{}ojasiewicza 11, 30-348 Krak\'ow, Poland}
}
\date{\today}

\begin{abstract}
Double electron ionisation process occurs when an intense laser pulse interacts with atoms or molecules. Exact {\it ab initio} numerical simulation of  such a situation is extremely computer resources demanding, thus often one is forced to apply reduced dimensionality models to get insight into the physics of the process.	
The performance of several algorithms for simulating double electron ionization by strong  femtosecond laser pulses are studied. The obtained ionization yields and the momentum distributions of the released electrons are compared, 
and the  effects of the  model dimensionality  on the ionization dynamics discussed.
\end{abstract}

\maketitle


	\section{\label{sec:introduction}Introduction}

A number of noteworthy phenomena occur during interactions of atoms and molecules with short strong laser pulses, for example  High order Harmonic Generation (HHG) and Nonsequential Double Ionization (NSDI). As experiments in strong field physics become more refined there is an increasing demand for performing \textit{ab initio} simulations that could provide an insight into observed phenomena. Several methods have been developed in that respect over the last twenty years. However, quite surprisingly, it is rather difficult
to find a  comparison between various proposed approaches. This is the primary aim of the present contribution. We shall concentrate on ionization of He atom as the simplest 
two electron system. 

Ideally the experimental data are to be simulated with full-dimensional theoretical models \cite{parker1998intense,parker2000time}. However, a full scale simulation requires huge computational resources. Practically, for low laser frequencies, it is doable only for systems within a single-active-electron approximation \cite{kulander1992time}. Thus, already two-electron correlation effects are almost beyond the scope of possibilities of the full-dimensional quantum treatment -- the only calculations performed in such a way 
in \cite{parker1998intense,
feist2008nonsequential,djiokap2017kinematical} are very hard to implement technically. These calculations, treating double-electron effects,  were performed exclusively for laser pulses with a carrier frequency in the UV spectral range, while most experiments deal with infrared and mid-infrared ranges, which makes applications of the full grid-based methods limited. Realistically, one is limited to design and use 
unsophisticated models with a reduced number 
of dimensions. The hardest part is to 
judiciously choose a coordinate system that, on one hand, 
captures ``all the important physics'' of the treated phenomena, and, on the other, is reasonably tractable with  available hardware and software resources. Finally, one has to admit that these low-dimensional models inevitably lack the precision and the predictive power of full-scale simulations.    

Nevertheless, most of the effects and tendencies observed in double ionisation are surprisingly well described, at least qualitatively, by models with reduced dimensionality \cite{becker2012theories}. This concerns,  in particular, NSDI, in which two electrons, after an initial excitation of one of them, share the excitation energy and ionize together.  Among the most prominent features, the ``knee'' structure in the double ionization yields, found in numerous experiments \cite{Walker94,fittinghoff1992observation}, was successfully simulated within the quantum mechanical 2D models \cite{lein2000intense,ruizPRA03,prauzner2007time,prauzner2008quantum} as well as with its classical analogues \cite{panfili2001comparing,panfili2002slow}. The ``finger-like'' structure of released electron momenta, obtained experimentally in \cite{weber2000correlated,rudenko2007correlated,kubel2014non,camus2012attosecond}, was reproduced with quantum \cite{staudte2007binary,
prauzner2007time} 
and classical \cite{ye2008classical} simulations. 
At the same time, other results, such as 
the theoretical prediction of the second plateau in high harmonic generation spectra \cite{koval2007nonsequential} 
have not been verified experimentally yet (to our knowledge) mainly because of the presently inaccessibly high precision needed for
their investigation.  All the reduced dimensionality models employ a linearly polarized field, as a system affected by an elliptically polarized field  necessarily needs the full dimensionality studies.

The first (and the simplest) restricted dimensionally two-electron model implemented is the so called ``aligned electrons'', or Rochester model \cite{bauer1997two}. With the nuclear motion not taken into an account, the $3+3=6$ dimensions of the problem are reduced to two, each one corresponding to the $z$-coordinate of an electron, where $z$ is collinear with the electric field direction. This model allowed one to observe the well pronounced ``knee'' structure \cite{lein2000intense}, as well as the momenta distribution \cite{chen2018rabi}. The most striking disadvantage of the method is due to  the distortions in the double ionization data: electrons are not allowed to escape with similar momenta because of the dominant effect of the Coulomb repulsion in the restricted geometry. 

This problem of the aligned-electron model has been eliminated in the Eckhardt-Sacha model by spatially separating the two axes \cite{sacha2001pathways}. Their direction is determined by the lines drawn by saddles of the field when its amplitude is
varied. Those lines 
are at the  angle of $\pm\pi /6$ with respect to the $z$ axis. While necessarily the reduced dimensionality has obvious
drawbacks (e.g. considering the motion along the saddle lines only neglecting possible 
 optimal paths across the saddles in full space) this model overcomes the obvious drawbacks of the aligned electrons model.
As a result, more reliable data for both ionization yield and momenta distribution were obtained for 
Eckhardt-Sacha model by Prauzner-Bechcicki {\it et al.} (ES approach) \cite{prauzner2007time,prauzner2008quantum,eckhardt2010phase}.

Another approach was proposed by Ruiz  {\it et al.} \cite{ruiz2006ab},
by making reasonable assumptions about the motion of electrons in the laser field. In the Center-of-Mass (CM) representation and with linear polarization for the laser light  one may assume that CM moves along the polarization axis. The resulting model yielded a striking success in reproducing the experimentally obtained parallel momenta distribution (momenta parallel to $z$ axis) \cite{staudte2007binary}.

All these geometries, the aligned electron, ES, CM or even the full dimensionality models, can be implemented in classical simulations. The key idea 
is to mimic the evolution of electronic systems in terms of classical trajectories that are sampled from 
an initial phase space distribution and are governed by classical Hamiltonian dynamics \cite{Leopold79}. Two groups of methods 
can be be distinguished: one with initial distributions calculated classically \cite{Leopold79,grochmalicki1991stabilization,Gajda92,Rzazewski94,Wojcik95,ho2007argon}, and one 
accounting for below-barrier tunneling for initial distribution calculation \cite{chen2000interpretation}. The first group 
is the most instructive one for a comparison of the data with corresponding quantum-mechanical computations. 
Considerable progress was made with these models for the study of ionization yields 
\cite{ho2005nonsequential,mauger2009strong} and momenta distributions \cite{panfili2001comparing}. 
On the other hand getting HHG spectra with classical means only seems impossible (see, however, \cite{berman2018coherent}). In particular the extend of the plateau depends on the quantum ionisation energy.

Generally, one can take some trusted analytical expressions as a reference. An important milestone in theoretical studies of atomic ionization was the creation of the Ammosov-Delone-Krainov (ADK) formula \cite{ammosov1986tunnel} that provides an extremely simple expression for the single tunneling ionization probability. It proved its reliability in comparison with numerous experimental data \cite{Walker94,fittinghoff1992observation,kondo1993wavelength,larochelle1998non}. So far, it can be used for comparison of the single ionization yield in different models. It is important to note, however, that the ADK rates are usually multiplied by an artificial constant factor to fit measured curves, thus one should be careful when treating absolute magnitudes.

Here we aim to compare the performance of the different computational models in determining the following experimental observables: yield of single ionization, of double ionization, and momenta distribution of the released electrons. 
We will focus our attention on ES and CM models, as they seem to be the
most advanced. The aim is to see how well they are able to
reproduce the experimental features qualitatively: we cannot expect that
reduced dimensionality models of this or any other type can be in quantitative
agreement with  experiments or truly three-dimensional simulations because of the vastly different
phase space volumes and ratios. Nevertheless, as we will have to rely
on low-dimensional models for some time to come, it will be interesting to
see how they compare - a study that, to the best of our knowledge, has not been
undertaken so far.

In Section \ref{sec:algorithms} we describe the algorithms used. In section \ref{sec:IY} we present the comparison of ionization yields and explain the observed differences. Section \ref{sec:moments} deals with comparison of electronic momenta. Conclusions are given
in Section\ref{sec:co}.  {Let us stress that our aim is not to find the model yielding ``most accurate'', i.e. in the closest agreement to the experiment, prediction. This is not 
possible for reduced dimensionality models. The models may serve to qualitatively describe experimental observations only. Still it is interesting to compare even the qualitative predictions different models yield. We are not aware of any such an earlier comparison.}
Throughout the text we use atomic units, unless stated otherwise.

	\section{\label{sec:algorithms}Description of the algorithm}
	
	\subsection{General remarks}

The quantum algorithms for both models are based on the operator splitting method. In short, once one has a Hamiltonian for the problem under consideration in the form of $H=H_1+H_2$, the solution of the corresponding time dependent Schr\"odinger equation (TDSE) for a wavefunction $\Psi(t)$ for a small time interval $\Delta t$ can be expressed according to the Suzuki-Trotter decomposition  as
\begin{equation}
\begin{split}
\Psi (t_0 + \Delta t) \sim
\exp\left( -i H_1 \frac{\Delta t}{2} \right) 
\cdot \exp\left( -i H_2 \Delta t \right) \cdot \\
 \exp\left( -i H_1 \frac{\Delta t}{2} \right)
\Psi (t_0).
\end{split}
\label{SOM}
\end{equation}
The wavefunction $\Psi (t_0 + \Delta t)$ is obtained by sequentially propagating the initial wavefunction $\Psi(t_0)$
(from right to left)
with Hamiltonian $H_1$ for a time $\Delta t/2$, 
then with $H_2$ for a time $\Delta t$ and then again with $H_1$ for a time $\Delta t/2$.
For a larger number of terms in Hamiltonian one can expand the formula in a straightforward manner. For instance,
if $H_2=H_3+H_4$ then one can write
\begin{multline}
\exp\left( -i H_2 \Delta t \right) \sim
\exp\left( -i H_3 \frac{\Delta t}{2} \right) 
\cdot \\ \exp\left( -i H_4 \Delta t \right) \cdot 
\exp\left( -i H_3 \frac{\Delta t}{2} \right)
\label{SOM2}
\end{multline}
and propagate sequentially as in the previous case.

Eigenstates of the system of interest are obtained by imaginary time propagation of a proper Hamiltonian with a Gaussian wavepacket as an initial state. Singularities in Coulomb-type potential terms are removed by replacing $1/x\rightarrow 1/\sqrt{x^2+\epsilon^2}$ (softening of the potential is applied in both real and imaginary time evolution). A soft-core parameter $\epsilon^2$ is chosen to align the ground state energy with the value of the full quantum problem at hand, in our case the ground state of He. {That assures the ionization potential of different compared models to be similar. Small changes of $\epsilon^2$ in quantum calculations only qualitatively affect the results obtained.}  In classical simulations the soft-core parameter plays the same role as in quantum simulations, but in this case, in order to ``stabilize'' the atom and prevent the autoionization process to take place, it must belong to the certain range of values that are usually higher than for the quantum models.

In the following we consider a linearly polarized laser pulse, described by an electric field component of the form
\begin{equation}
F(t)=F_0f(t)\sin(\omega t + \phi),
\end{equation}
where $F_0$, $f(t)$, $\omega$ and $\phi$ are the field amplitude, the time-dependent envelope, the frequency and the carrier-envelope phase of the pulse, respectively. For comparison of different models we take a laser pulse with the frequency $\omega=0.06$ 
(corresponding roughly to the wavelength of 800 nm), the phase $\phi=0$ and the sine-squared envelope:
\begin{equation}
f(t)=\sin^2\left(\frac{\pi t}{T}\right),
\end{equation}
where $T={2\pi n}/{\omega}$ is the duration of an $n$-cycle pulse. Here, we take $n=4$ as a typical value.

To prevent a nonphysical reflection of the wavefunction from boundaries of the numerical grid, absorbing boundary conditions are applied, i.e. starting from a fixed distance from the edge of integration region,  the wavefunction is multiplied by a function 
that smoothly decreases in the direction of boundaries.

	\subsection{Algorithm for Eckhardt-Sacha model}

The Eckhardt-Sacha model \cite{sacha2001pathways} assumes that the ionization of atoms occurs mainly along two directions 
$\textbf{r}_1$ and $\textbf{r}_2$, forming $\pi/6$ angle with the $z$ axis, due to the location of saddle points of the energy surface. The resulting 2D Hamiltonian, in  the length gauge, may be written as:
\begin{multline}
H = \sum_{i=1}^{2} \left( \frac{p_i^2}{2} - \frac{2}{\sqrt{r_i^2 +\epsilon^2}} +  \frac{F(t)\sqrt{3}}{2}r_i \right) + \\ \frac{1}{\sqrt{(r_1-r_2)^2 + r_1 r_2 + \epsilon^2}},
\label{eq:1plus1q}
\end{multline}
%
where $F(t)$ is the electric field value, $r_i$ and $p_i$ are the position and momenta operators for both electrons, 
while $\epsilon^2$ is the parameter introduced to {soften} the Coulomb singularity in the reduced dimensionality model. 

In order to obtain the wavefunction evolution in time one splits the Hamiltonian into kinetic and potential parts and uses Eq.~(\ref{SOM}). It is worth noting that the evolution of the kinetic part is efficiently done in momentum space, as it reduces to a simple multiplication, 
whereas the evolution of the potential part is best computed in coordinate representation \cite{prauzner2007time}. This strategy allows one to eliminate the numerical differentiation and thus to  increase the precision of calculations. The transformation between these representations is realized via fast Fourier transform (FFT) routines.

\begin{figure}
\includegraphics[width=1.0\linewidth]{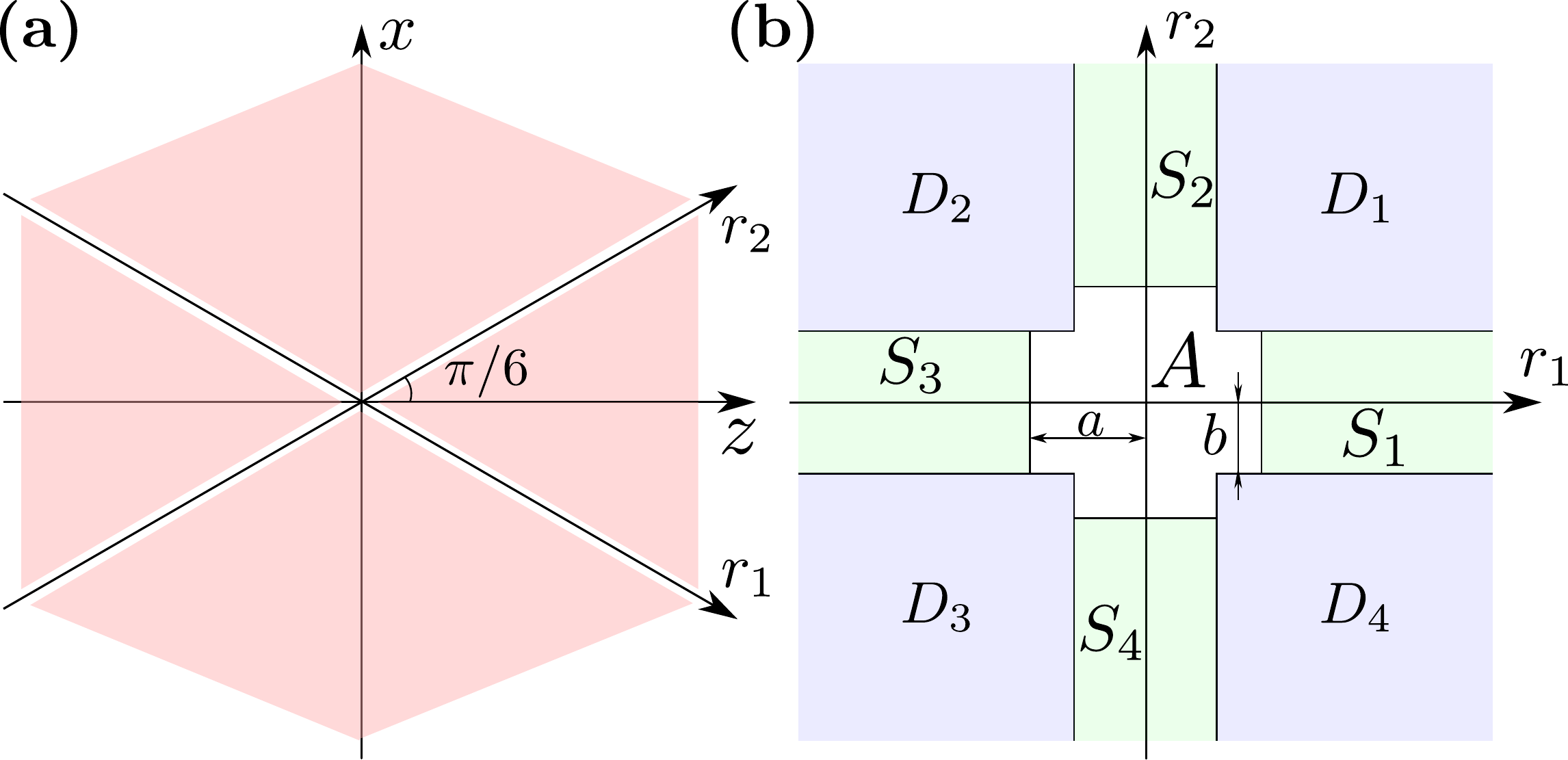}
\caption{Geometry of the ES model. (a) The saddle tracks forming the directions along which electrons are allowed to move. The polarization axis points along $z$. (b) Different regions of the configuration space used to define the state of the system: the neutral atom $A$, singly charged ions $S_i$, and doubly charged ions $D_i$. The parameters $a=12.5$ a.u. and $b=6$ a.u.}.
\label{geometry_PB}
\end{figure}

Ionization yields are calculated by integrating probability fluxes through borders between different spatial areas corresponding to a stable atom ($A$) or a single ($S_i$) and a double ($D_i$) ionization event, as depicted in Fig. \ref{geometry_PB}. The method, originally proposed by Dundas {\it {\it et al.}.} \cite{dundas99}, was extended in \cite{prauzner2007time,prauzner2008quantum} in a way that allowed to distinguish between direct and indirect double ionization events. That is accomplished thanks to different values of 
parameters $a$ and $b$, defining different regions in the configuration 
space. In particular, for $b < a$ simultaneous double ionization 
events can be detected as transitions across the common border between neutral atom,
$A$ and regions $D_i$ that correspond to double ionization. For $b = a$ 
those border shrinks to a point, and disappear for $b>a$, requiring more
complex indicators for the direct double ionization.

Of course, the choice of $a$ and $b$ affects the results, but fortunately
only weakly so. For instance, making them twice bigger, 
will change quantitative values for e.g. ionization yields, but does
not, as verified by us, change the qualitative picture, and will therefore
still allow us to compare the reduced dimensionality models. The actual
values used in our simulations are the same as 
those taken in previous studies \cite{dundas99,prauzner2007time} for double ionisation studies.
 
Within the same model it is also possible to obtain momenta distributions, however, a different computational approach is needed \cite{lein2000intense,prauzner2008quantum}. First, one needs to rewrite the Hamiltonian~(\ref{eq:1plus1q}) in the velocity gauge, where the vector potential is given by
\begin{equation}
A(t)=-\int_0^t F(\tau){\rm d}\tau.
\end{equation}
Next, it is assumed that electrons that travel a large distance from the nucleus, say 200 a.u. and more, experience a negligible Coulomb interaction with the nucleus, are unlikely to turn back and follow an evolution governed predominantly by the laser field. For such electrons it is then plausible to assume that all Coulomb terms may be ignored, leaving only the kinetic part in the Hamiltonian (recall that now the velocity gauge is used). In such a case the evolution is efficiently performed in the momentum representation as it reduces to a multiplication by a proper phase factor. Furthermore, evolving the wavefunction in the momentum representation allows one to keep all information about an infinite position space -- no parts of the wavefunction are lost due to the absorbing boundary conditions. The ES model describes a two-electron system, thus it is necessary to consider also an intermediate case, i.e. when only one electron is far away and the other is still relatively close to the 
nucleus. In such a case, only the interaction of the distant electron with the nucleus and the other electron are neglected, while the electron close to the core is evolved with Coulomb electron-nucleus interaction included. Eventually, the full evolution is performed in three different regions, i.e. in a region with both electrons close to the nucleus (evolution with the full Hamiltonian), in a region with one electron close to the nucleus and the other at a larger distance (semi-approximate Hamiltonian), and in a region with two distant electrons (full-approximate Hamiltonian). At the end of the calculations, one collects all parts of the wavefunctions from the three regions in the momentum representation, while the part related to bound states is smoothly extracted. The squared modulus of the wavefunction in the momentum space gives a momentum distribution. A detailed description of the method, including the procedure for 
transferring the wavefunction between different regions, is presented in \cite{prauzner2008quantum}.

For standard simulations within the ES model the following parameters are used: a soft-core parameter of $\epsilon^2=0.6$ 
(yielding the correct ground state energy for He), the spatial grid step $\Delta r_1 = \Delta r_2 =  0.2$, 
and the  temporal step of $\Delta t = 0.05$.

The computations for an aligned electrons model employ the same algorithm in this work. The change of the system geometry is expressed in the change of the Hamiltonian (\ref{eq:1plus1q}):
\begin{multline}
	H = \sum_{i=1}^{2} \left( \frac{p_i^2}{2} - \frac{2}{\sqrt{r_i^2 +\epsilon^2}} + F(t)\cdot r_i \right) + \\ \frac{1}{\sqrt{(r_1-r_2)^2  + \epsilon^2}}.
	\label{eq:aligne}
\end{multline}

	\subsection{Algorithm for Center-of-Mass model}

The CM model follows from the simple observations made by
 Ruiz {\it et al.} \cite{ruiz2006ab}, that classically, in linearly polarized fields, the component of the atomic CM momentum
perpendicular to the field polarization direction is conserved, and thus may be put to zero. 
For two-electron systems, one of the coordinates then vanishes. The conservation of the angular momentum projection on the polarization $z$ axis reduces the problem further, so that only three dimensions out of the initial six remain. 
They are the relative cylindrical radial coordinate $\rho = |\rho_1 - \rho_2|$ and the field-parallel coordinates $z_1$ and $z_2$
of both electrons, or, equivalently, $Z_c=(z_1 + z_2)/2$ and $z = (z_1 - z_2)$ for the coordinates of the CM and the relative position. 
The last coordinate pair is useful as the $z$ coordinate decouples from the electric field. 
The momenta operators in CM representation have the form 
as $\textbf{P} = \textbf{p}_1 + \textbf{p}_2$ and $\textbf{p} = (\textbf{p}_1 - \textbf{p}_2)/2$. 
The Hamiltonian of the system is then
\begin{equation}
\label{eq:cmq}
H = H_{Z_c} + H_z + H_{\rho},
\end{equation}
where
\begin{equation}
\begin{split}
& H_{Z_c} = -\frac{1}{4} \frac{\partial^2}{\partial Z_c^2} + A_Z^2 -iA_Z \frac{\partial}{\partial Z_c} + \frac{1}{3} V \\
& H_z = -\frac{\partial^2}{\partial z^2}  + \frac{1}{3} V \\
& H_{\rho} = -\frac{1}{\rho}\frac{\partial}{\partial \rho} \left( \rho \frac{\partial}{\partial \rho} \right)  + \frac{1}{3} V ,
\end{split}
\label{parH}
\end{equation}
with $A_Z$ the vector potential of the laser field (here the velocity gauge is used, although the length gauge may be used as well), 
and $V$ is the sum of 
Coulombic interactions:
\begin{multline}
V = \frac{1}{\sqrt{z^2 + \rho^2}} - \frac{2}{\sqrt{(Z_c - \frac{z}{2})^2 +  \frac{\rho^2}{4} + \epsilon^2}} \\ - \frac{2}{\sqrt{(Z_c + \frac{z}{2})^2 + \frac{\rho^2}{4} +\epsilon^2 }} .
\end{multline}

\begin{figure} 
\includegraphics[width=0.8\linewidth]{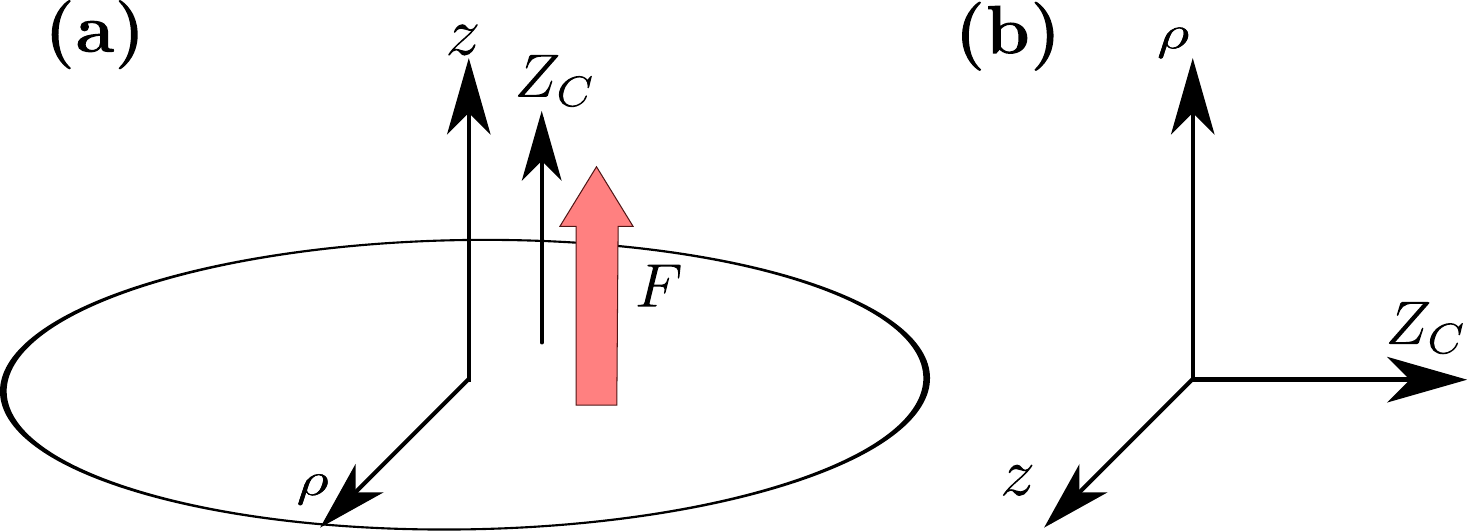}
\caption{ (a) Geometry of the CM model. $Z_c$ and $z$  -- the directions of Center of Mass and the relative electronic coordinate, respectively,  are collinear with the electric field polarization direction. $\rho$ is the cylindrical radial direction of relative motion in the $x-y$ plane. (b) Dimensions for evaluating the problem. }
\label{geometry_CM}
\end{figure}

The TDSE is solved via the operator splitting technique as described above, see Eqs.~\eqref{SOM}-\eqref{SOM2} for Hamiltonian (\ref{eq:cmq}). Note the unusual symmetric distribution of the potential 
onto all parts 
in Eq.~(\ref{parH}). 
Elementary unitary steps in Eqs.~\eqref{SOM}-\eqref{SOM2} are  evaluated with the Crank-Nicolson method. It is particularly efficient for this problem as in each unitary step only derivatives along a single coordinate  are present, resulting in tridiagonal matrices that may be efficiently inverted. The wavefunction is defined, of course, on a 3-dimensional grid. More details on the algorithm may be found elsewhere \cite{alexisthesis}.

Ionization yields are computed by integrating the wavefunction in the region $r>12.5$ a.u.; for calculating the double ionization yield, 
the additional conditions used are $\sqrt{z_1^2+\rho^2/4} > z_0$ and $\sqrt{z_2^2+\rho^2/4} > z_0$ as well as $z_0=12.5$ a.u. Again, as for SE model, the choice of $z_0$ is to some extend arbitrary so we take $z_0=a$ (the original proposition of the model \cite{ruiz2006ab} uses $z_0=12$ a.u.).

For calculating the final momenta distribution, the transformation of the wavefunction corresponding to the double-ionized state from the coordinate representation to the momentum representation is realized by a fast Fourier transform along the $z$ and $Z_c$ directions and by a Hankel transform along the $\rho$ direction. Such a wavefunction is obtained by extracting bound states from the full wavefunction. In our algorithm the spatial criterion for separating bound and free states is used. The part of the wavefunction corresponding to the region with radius less than the given value $z_0$ is multiplied by a function that tends to zero in the neighborhood of zero radius and becomes equal to unity when the radius equals  $z_0$. We use a Gaussian function for this purpose, with standard deviation equal to 10 a.u. The value of $z_0$ for calculating momenta distribution is taken to be 30 a.u.

Both the ionization yields and the momenta distributions obtained using our implementation  are in agreement with those obtained by authors of the original algorithm \cite{chen2010double}.

For standard simulations the following parameters are used: a soft-core parameter of $\epsilon^2=0.135$, spatial grid step $\Delta z = \Delta Z_c = \Delta \rho = 0.3$ a.u. and temporal one $\Delta t = 0.05$ a.u. 

	\subsection{Algorithms for classical simulations}

One of the important advantages of classical algorithms in comparison to quantum ones are their higher efficiency and lower requirements for computational resources.
They allow one to analyze the original two-electron system without dimensional constraints and for a wide range of field intensities.
Due to the classical consideration of electrons motion one can trace back the particular double or single-ionized trajectory and can deduce additional information about the mechanism of ionization and the preliminary conditions of the system that favor an ionization event.

The classical algorithm for our two-electron system is based on the analysis of the Hamiltonian for the respective model 
by numerical integration of the canonical equations of motions. As the system which we consider is non-integrable, 
special attention should be paid to stability of the numerical algorithm used for their solution \cite{hairer2006geometric,efimov2014analysis}.
In our work the numerical integration was based on the symplectic Runge--Kutta--Nystr\"{o}m algorithm \cite{blanesJCAM02} with its parameters chosen properly to give a minimal effective error.

In the most general setting, the studied two-electron system is described by a 6D Hamiltonian of the following form:
\begin{multline}
\label{eq:hamil3d}
H=\sum_{i=1}^{2}\left( \frac{\textbf{p}_{i}^{2}}{2}-\frac{2}{\sqrt{\textbf{r}_{i}^{2} + \epsilon^2}}+\textbf{F}(t) \cdot \textbf{r}_{i} \right) \\ +\frac{1}{\sqrt{(\textbf{r}_{1}-\textbf{r}_{2})^{2}+\epsilon^2}},
\end{multline}
where $\textbf{p}_{i}=\{ p_{x_i}, p_{y_i} , p_{z_i} \}$ and $\textbf{r}_{i}=\{ x_i, y_i , z_i \}$ are 3D momenta and position vectors of  electrons $i=1$ and $2$, respectively.

The initial phase space coordinates required for integration of the canonical equations of motion are generated by the pilot atom's two-electron trajectory technique at zero field amplitude \cite{panfiliOE01, panfiliPRL02}. The pilot trajectory is started at zero position space coordinates, whereas its initial momenta are obtained by random distribution of residual energy $E_{r}$, which is the difference between the ground state energy and a fictitious potential energy arising from including the smoothing factor $\epsilon$, i.e. $E_{r}=E_{g}-1/ \epsilon$. The pilot atom's two-electron trajectory is integrated until the energetically allowed position and momenta spaces are fully populated.

To reach the entire population of the energetically allowed space for our case, the pilot atom's trajectory was run for time $t=10^{4}$ a.u., producing an ensemble of about $10^{7}$ initial points. The initial 
coordinates for the pilot atom's two-electron trajectory are chosen randomly.
One should note that some alternative techniques for generating initial conditions are possible. An important example is a widely used microcanonical ensemble technique \cite{Leopold79,Gajda92,Wojcik95,maugerJPB09}. We implemented and tested both methods and have found that the choice between them is irrelevant for generating initial distribution for our problem as both lead to a very similar value of  ionization yields. We use the pilot atom's two-electron trajectory technique as it provides much faster calculations. The ground state energy is set to $E_{g}=-2.936$ a.u. for all our classical simulations. 

In order to obtain reasonable information about ionization yield a large ensemble of $10^7$ trajectories is used for calculations in the presence of a laser field. At the end of the pulse, single and double ionization events are extracted by applying the spatial criterion: the electron is considered to be ionized if its distance from the nucleus is large, i.e. $r>100$ a.u.

In contrast to atoms described in quantum formalism, classical atoms may experience autoionization in absence of external field due to intensive many-body Coulombic interactions \cite{haan1994numerical,mauger2010recollisions}. Such an autoionization may be several orders more intensive than the expected ionization by external laser field, thus one has to eliminate the effect. It is done by introducing the $\epsilon$ term to the Hamiltonian in the same manner as in the quantum mechanical models. For all our classical simulation we set $\epsilon^{2}=0.6$. 

Following the method described above the numerical simulations for quantum ES, aligned electron and CM models given by Eqs. (\ref{eq:1plus1q}), (\ref{eq:aligne}) and (\ref{eq:cmq}), respectively, are reproduced classically in the framework of 
Hamiltonian dynamics.

	\section{\label{sec:IY}Ionization yields comparison}

\begin{figure}
	\includegraphics[width=1.0\linewidth]{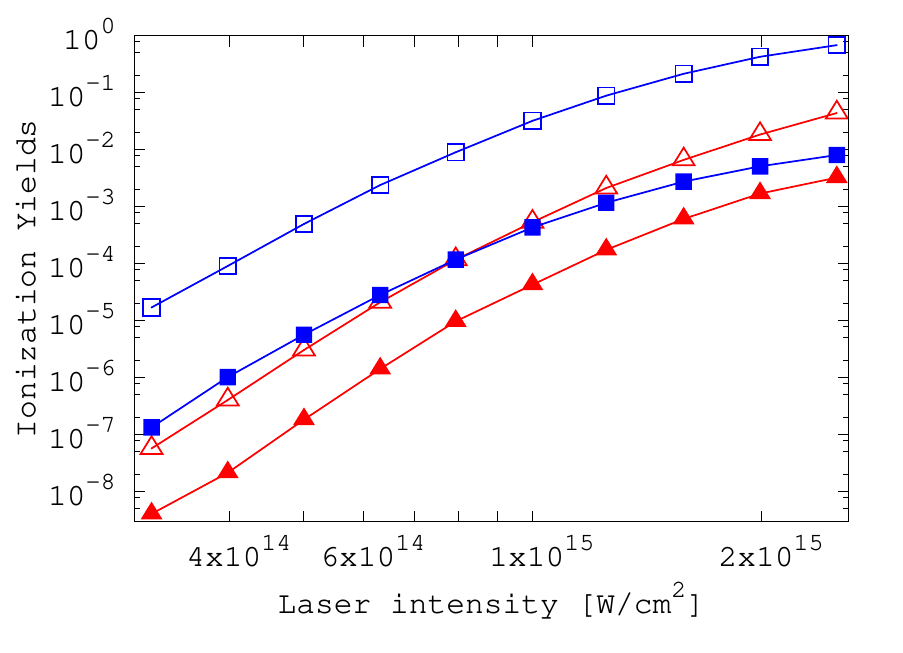}
	\caption{Single and Double ionization yields (probability of a given process) as a function of  laser field intensity obtained by quantum-mechanical simulations. Single and double ionization yields are marked by empty and filled symbols, respectively; results for the ES model are marked by blue squares, for the CM model -- red triangles. In both cases a 4 cycle, ``$\sin^2$'' pulse was assumed.}
	\label{IY_quantum}
\end{figure}

\begin{figure}
	\includegraphics[width=1.0\linewidth]{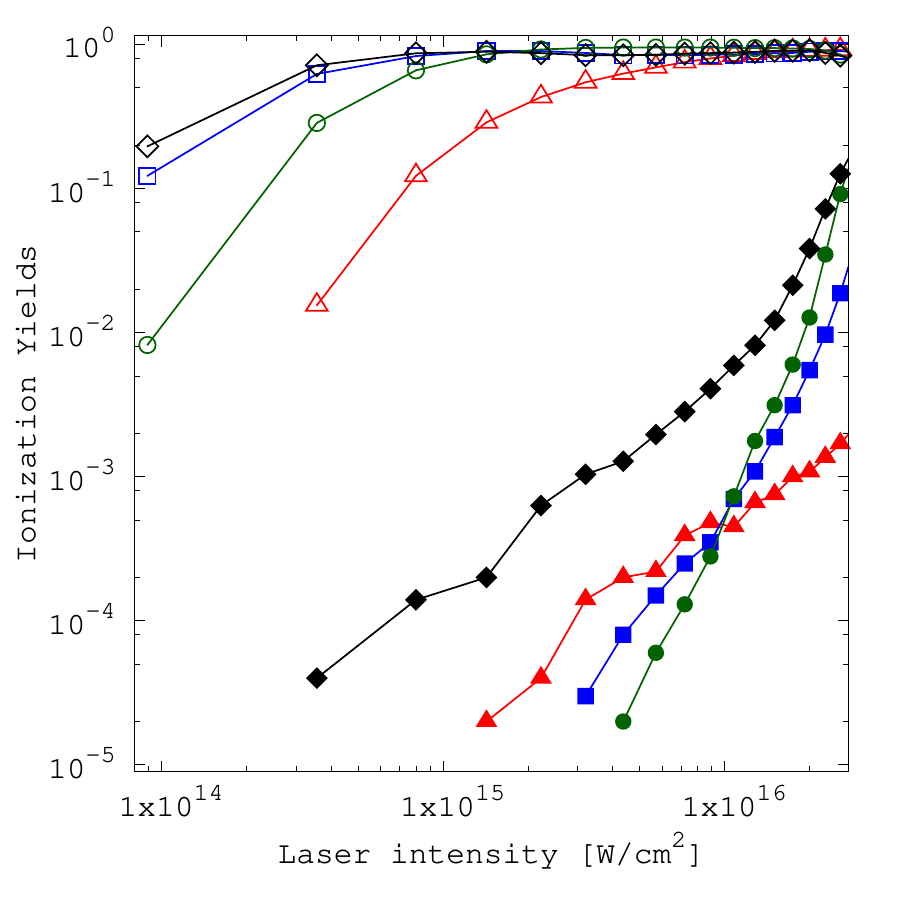}
	\caption{ Single and Double ionization yields as a function of the laser field intensity obtained by classical simulations. Single and double ionization yields are marked by empty and filled symbols, respectively; results for the ES model are marked by blue squares, for the CM model -- red triangles, Rochester model -- black diamonds, full 3D classical model -- green circles. For all the
		simulations 800 nm 4 cycle pulses of ``$\sin^2$'' shape were used.}
	\label{IY_classical}
\end{figure}

The Ionization Yield (IY) is one of the most important quantities characterizing ionization dynamics of atomic and molecular systems. In Fig.~\ref{IY_quantum} and Fig.~\ref{IY_classical} one can see the dependence of the IY  on the laser field intensity 
obtained from
quantum and classical simulations. 

The first striking observation coming from the analysis of both graphs is the low probability of ionization in the CM model -- for both classical and quantum calculations the yield appears to be less than 
that for 1D+1D and for (classical only) 3D+3D models. This is quite understandable 
from the 
classical mechanics. So, let us first explain the difference between CM and aligned electron models, based on geometrical arguments.

The condition of setting the CM radial coordinate to zero is in fact a kind of a holonomic constraint. It implies the rule that both electron radial coordinates are the same, up to a sign: $\rho_1=-\rho_2$. For single ionization the electron's escape is most probable along the 
$z$ polarization axis, or $\rho_1 \sim \rho_2 \sim 0$. While for the full 6D case this condition by no means restricts the second electron position, in the CM case the second electron should have the same radial coordinate. In other words, the number of possible electronic spatial configurations that lead to an escape of an electron is considerably smaller for the CM model, resulting in a smaller ionization output.

The same arguments help to understand the CM vs. 1D+1D comparison. While for the CM case the configurations with $\rho_1 \sim \rho_2 \sim 0$ constitute a minority, for the aligned electrons case the configurations with $\rho_1 = \rho_2 = 0$ are the only possible ones
 yielding a much larger output. Although the ES 1D+1D model puts both axes in a nonzero angle to the $z$ axis, in fact it affects only the effective field strength value, multiplying it by a cosine of this angle, thus allowing one to use the same explanation.

The Sequential Double Ionization (SDI) yields  follow the same scenario: the second electron ionization implies that  both electrons are in the neighborhood of the $z$ axis, thus reducing the number of possible configurations and consequently the total yield. At the same time the NSDI signal requires the electrons correlation which occurs only when both electrons are close to each other ($\rho_1 - \rho_2 \sim 0$), but once $\rho_1=-\rho_2$, it leads to the pronounced condition $\rho_1 \sim \rho_2 \sim 0$ and thus a lowered ionization yield.

Since all the above explanations are essentially geometrical, they can be applied to the quantum case as well: the quantum 
data show a similar trend.

\begin{figure}
	\includegraphics[width=1.0\linewidth]{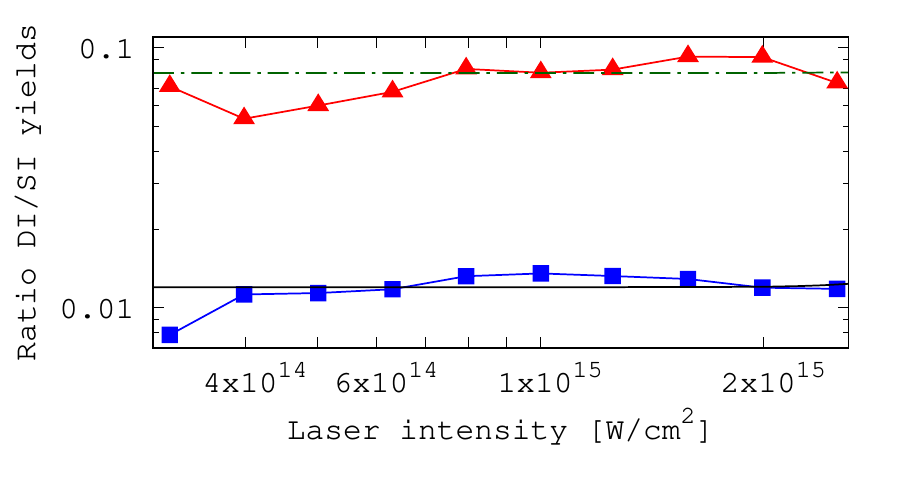}
	\caption{Ratio of double to single ionization yields as a function of the laser field intensity for the CM (triangles) and the ES (squares) models  obtained by quantum simulations. The solid and dash-doted lines denotes the ratio calculated with the rate equations with constants $C=0.012$ and $C=0.08$ correspondingly.}
	\label{IY_ratio}
\end{figure}

The important parameter to look at is the ratio between yield for
doubly and singly ionized atoms. Its dependence on field intensity
is depicted in Fig.~\ref{IY_ratio}.  The ratio between
double and single ionization yields for both algorithms has
approximately the same ``flat'' shape in the regime of
intensities studied. It correlates very well with the saturation
of that ratio observed in experiments \cite{Walker94} as well as in double
ionization model \cite{Bhardwaj01} based on rescattering mechanism \cite{Corkum93}.
On the other hand, such a ratio allows for a quantitative
comparison with experiment since volume averaging effects
(due to laser intensity variation across the sample) while
affecting strongly the yields themselves may modify this ratio
by a factor of two at most.
Both the experiment \cite{Walker94} and
theory \cite{Bhardwaj01,Yudin01,chen2018ratios} give the saturation value of He$^{2+}$/He$^+$
being of the order of $10^{-3}$, an order of magnitude smaller
than the ratio obtained for ES model, and almost two orders
smaller than this ratio in CM model (compare Fig.~\ref{IY_ratio}).
This
discrepancy again indicates that the
predictions of restricted dimensionality models may be qualitative
at best.
One could make an attempt to calculate this ratio classically
-- but Fig.~\ref{IY_classical}  reveals that classical ensemble calculations fail
to reproduce any plateau in the ionization yields ratio for
the chosen Coulomb smoothing parameter $\epsilon^2=0.6$. By allowing
for more freedom and using different smoothing parameters for
electron-nucleus and electron-electron interactions one may
significantly modify the ratio between doubly and singly
ionized species bringing it closer to experimental values.
We mention such a possibility only as we do not want to
complicate the studied model further.

It is not surprising
that ionization yields deduced from classical simulations are much bigger than those from quantum calculations. 
This seems to be a general property of all the classical simulations as seen from the 
comparison of data from classical \cite{ho2005nonsequential,ho2005classical,ho2006plane,mauger2009strong,mauger2010recollisions,mauger2010fromrecollisions} and quantum \cite{brics2014nonsequential,lein2000intense,ruizPRA03,prauzner2007time,prauzner2008quantum,parker2007single} 
simulations.

One should note that both quantum algorithms (SE and CM) have their limitations on the laser field intensities for which the calculations can be performed. Both algorithms become inadequate at obtaining ionization yields for electric field amplitude less than $10^{14}\text{W/cm}^2$ as any possible reduction of spatial and temporal grid size gives rise to parasitic numerical effects. The CM algorithm also has its upper limit on intensity: larger intensities and laser field wavelengths require larger spatial grid sizes and thus, a larger amount of computer memory -- as well as computation time. For 800 nm and an intensity of $2.5\cdot 10^{15}\text{W/cm}^2$ in the laser field the converged calculations require already 128 GB of memory. 

\begin{figure}
	\includegraphics[width=1.0\linewidth]{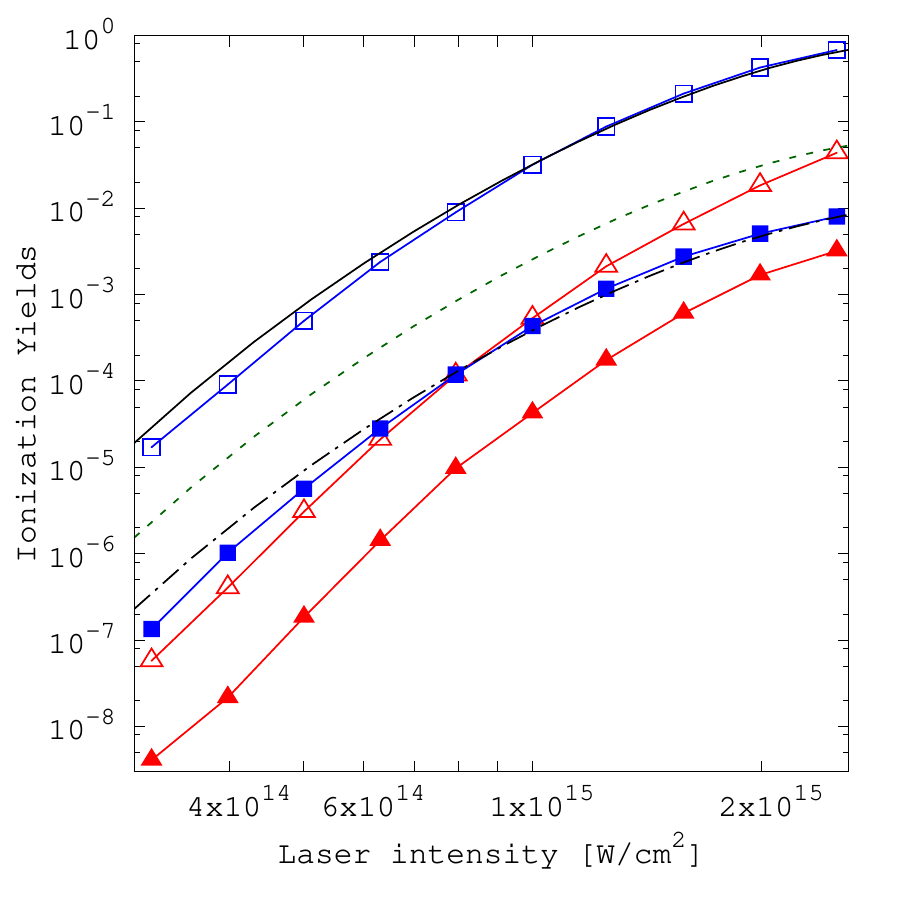}
	\caption{Single and Double ionization yields as a function of the laser field intensity. The points correspond to quantum-mechanical simulations as described in Fig. \ref{IY_quantum}.
		The lines without symbols correspond to yields obtained via Ammosov-Delone-Krainov formula. The solid line denotes Single Ionization yield. Dashed-dotted and dashed lines both correspond to Double ionization yields obtained with correct, $C=0.012$, and incorrect, $C=0.08$, coefficient,  correspondingly.}
	\label{ADK}
\end{figure}

The Ammosov-Delone-Krainov formula \cite{ammosov1986tunnel} was developed for calculating the rate of single electron ionization within the single active electron (SAE) approach under the influence of electric field $F(t)$ \cite{ilkov1992ionization}:
\begin{multline}
	W_{ij}^{\rm ADK} = \left( \frac{3e}{\pi} \right)^{\frac{3}{2}} \frac{j^2}{3n_{ij}^{*3}} \frac{1}{2n^{*}_{ij}-1} \cdot \\
	\left( \frac{4ej^{3}}{(2n_{ij}^{*} - 1) n_{ij}^{*3} |F| }  \right)^{2n_{ij}^{*} - \frac{3}{2}} \cdot \exp \left( -\frac{2j^3}{3n_{ij}^{*3} |F|}  \right),
\end{multline}
where $j\to i=(j-1)$ denotes ionization of atoms/ions of charge $i$, $n_{ij}^{*} = j/\sqrt{2E_{ij}}$ is an effective principal quantum number related to ionization energy $E_{ij} = E_0^{j+} - E_0^{(j-1)+}$. Such a rate can be used for setting the ionization rate equations \cite{huilier1983multiply} for populations of neutral atom $P_0$, single $P_1$ or double $P_2$ ionized ion:
\begin{equation}
	\begin{aligned}
	& \dot{P_0} = -W_{01}P_0 - W_{02}P_0 , \\
	& \dot{P_1} =  W_{01}P_0 - W_{12}P_1 , \\
	& \dot{P_2} =  W_{02}P_0 + W_{12}P_1.
	\end{aligned}
	\label{rate_eq}
\end{equation}
The rate $W_{02}$ of NSDI cannot be determined from the ADK approach, but for determining it we employ the fact that for the region of field amplitudes of interest the SDI is much less intensive than NSDI, 
and the ratio of $\text{He}^+$ and $\text{He}^{++}$ yields is almost constant. Thus one can put 
$W_{02}=W_{01}\cdot C$, where the constant $C=0.012$ is defined phenomenologically from TDSE-based ES numerical calculations (see Fig. \ref{IY_ratio}). The results of the straightforward integration of Eqs.~(\ref{rate_eq}) is shown in Fig. \ref{ADK}. 
The resulting curves are sensitive to value of $C$. To show that
we performed the integration of Eqs.~(\ref{rate_eq}) with an ``incorrect'' value of $C=0.08$ corresponding to data obtained by the CM algorithm; the resulting double ionization curve does not properly fit numerical data.

Surprisingly, the ionization yields obtained with the ADK formula coincide with the ones obtained using the ES algorithm very well. 
At the same time, ADK is known to fit the full 6D He calculation provided by Taylor {\it et al.} \cite{parker2007single}. Thus one can conclude that the ES algorithm provides data that may be qualitatively correct, which is quite a surprising and refreshing result. On the other hand, some of the agreement may be purely accidental.

	\section{\label{sec:moments}Photoelectron momenta distribution comparison}

\begin{figure*}
	\centering
	\includegraphics[width=\textwidth]{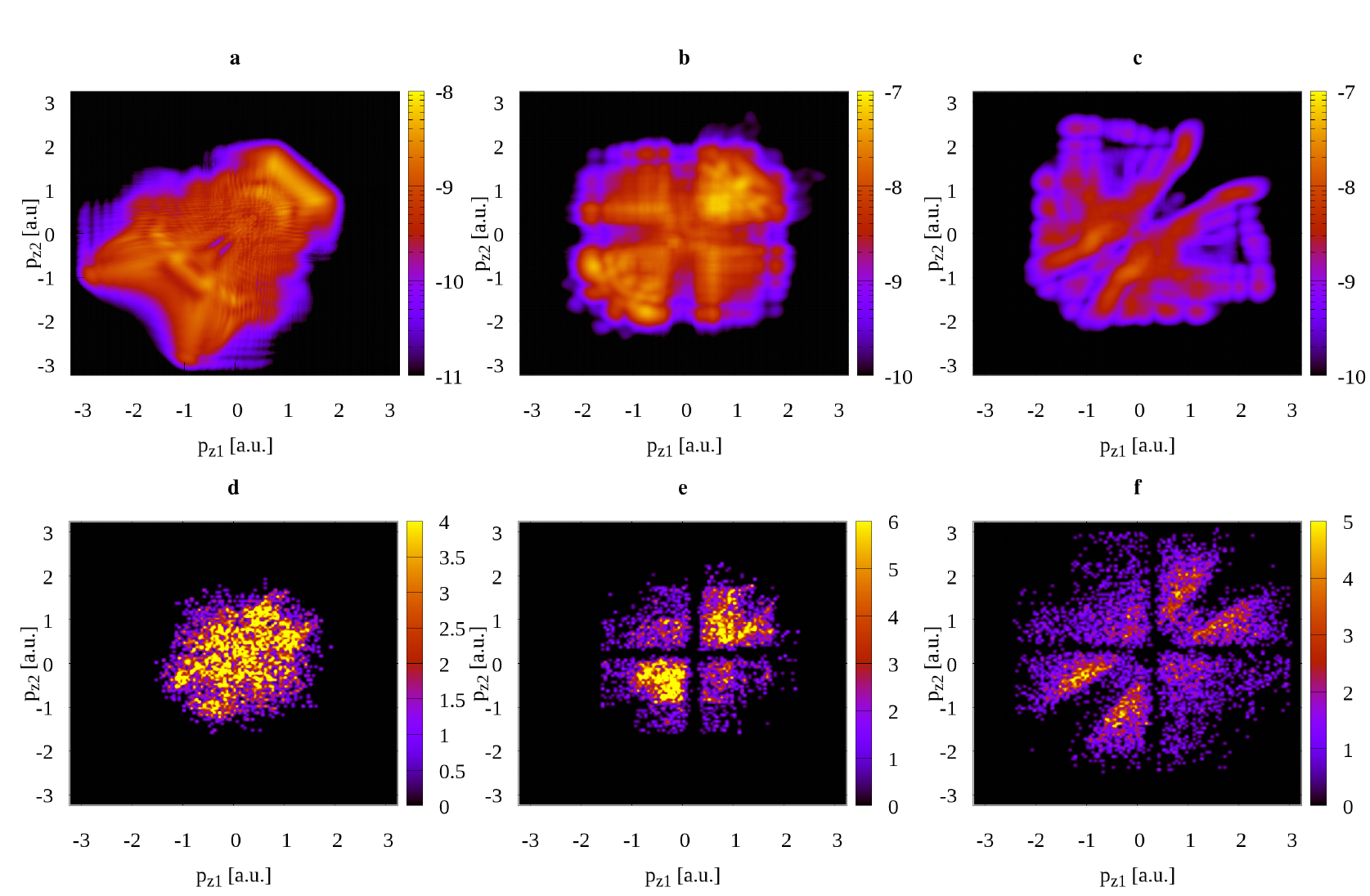} 
	\caption{
		Released electrons momenta distribution computed with quantum (a,b,c) and classical (d,e,f) simulations for the CM (a,d), the ES (b,e) and the aligned electron (c,f) models. Simulations were done for the 800 nm field of intensity of $6.3\cdot 10^{14}$ $ \text{W}/\text{cm}^2$. Parameters of grid for quantum models: for the ES and the aligned electron plot $4096 \times 4096$ points with coordinate step size 100/512 a.u.; for the CM model $3372\times 1686 \times 454$ points with coordinate step size 0.3 a.u.; laser pulse consists of 4 cycles. The plots (b,c) are smoothed with gaussian functions of 0.12 a.u. width. The value of $\epsilon^2 = 0.58$ in (c) which corresponds to the same energy of the ground state as in (b). Classical data obtained with use of $4\cdot 10^7$ sample trajectories for the CM model (d),   $4\cdot 10^7$ sample trajectories for the ES model (e) and $5\cdot 10^6$ sample trajectories for the aligned electron model (f). For classical simulations the same spatial criterion of double 
ionization event was used, 8-cycle  laser 
pulse was applied. }
	\label{Momenta}
\end{figure*}

A study of the momenta distribution is necessary for understanding the mechanism of double ionization \cite{staudte2007binary}. It constitutes a corner stone of the photoelectron holography technique \cite{huismans2011time}. ES, aligned electron and CM methods are able to provide parallel momenta distributions, i.e. a snapshot of electronic wave functions in momentum representation as a function of first and second electrons' momenta parallel to the $z$ electric field direction. The corresponding sample plots
are given in Fig. \ref{Momenta} (a,b).

The differences between these two figures can be explained by two factors: technical and geometrical. The first concerns the 
computation of the wavefunction in momenta representation. For the part of the wavefunction that reached the boarders of physical space, the ES algorithm preserves it in the momentum representation. Thus, the information about electrons that escape the computational space before the end of the evolution -- and thus presumably the fastest ones -- is not lost \cite{prauzner2008quantum}. Our realization of the CM algorithm does not possess such an ability, and the information about some portion of fast moving electrons may be missing provided the spatial grid for computations is not large enough.

The geometrical factor explains the presence and intensity of the interference picture. In the ES model the propagation of electronic waves is restricted to the plane, making their interference clear (corresponding image is not shown) \cite{prauzner2008quantum,shaaran2012causality}. For experimental relevance one should  introduce smoothing by Gaussian functions, this is done in Fig.~\ref{Momenta}(a,b).  Increasing the number of dimensions provides the electronic system with a much larger number of possible quantum paths leading to blurring of the interference structure. In this sense the CM model simulations yield more realistic predictions as experiments deal with full 3D problems.
The CM model is supposed to behave better than the ES one when multi-dimensional details of electron-core rescattering processes are needed,
eg. for
problems of photoelectron holography.

One should note that as the CM code is evaluated in the coordinate representation, for the sake of accounting of all the produced photoelectron momenta, one should keep the coordinate space rather large to not let any part of the wavefunction escape from it. Thus with the same set of parameters given, calculating momenta distributions requires  many times larger spatial grids 
than calculating ionization yields or high harmonics spectra. The ES algorithm does not have such a complication, as the part of the electronic wavefunction that moved far enough from the atomic core is then treated in the momentum representation, 
which requires a relatively small grid size.

On the other hand, the ES algorithm suffers from the problem of the ``empty cross'' in the momenta distribution plot \cite{prauzner2008quantum}. The cross occupies space along the $p_{r_1}$ and $p_{r_2}$ axes; the wavefunction inside it has a magnitude many orders smaller than in the neighbouring regions. It arises from cutting the wavefunction -- in coordinate representation -- in the area with coordinates close to zero. It is this area that contains most of the low-momenta electrons. Thus one should be careful with extracting the bounded part of the wavefunction. We have found that for the ES model the optimal value of distance at which the cutting is performed is 50 a.u. The CM algorithm is affected by this feature much less, mostly because electrons with zero momentum $p_z$ are not restricted to be in the close neighborhood of the atomic core, but can stay far from it due to the $\rho$ coordinate. 

The described factor is also geometrical, and can be illustrated with classical trajectory calculations. In Fig.~\ref{Momenta}(d,e) one can see  two plots for the $z$-axis momenta distribution of two electrons for the same problem studied.
First, one can note that the general shape of the distribution is quite similar to that of the quantum case in Fig.~\ref{Momenta}(a,b).
This leads to the conclusion that the classical calculations are much more appropriate in studies of electronic momenta distributions than for ionization yield studies. Second, the ``cross'' in the momenta distribution is present quite prominently 
in both the ES and aligned electron cases, and it arises for the same reason: electrons with $p_r \sim 0$ are located in the area $r \sim 0$. For this set of classical calculations the length of the laser pulse was taken to be twice larger than that for the quantum computations, as it is hard to collect enough double ionization events for a 4-cycle pulse.
	
One may observe, furthermore, that quantum momenta distributions for the same intensity are significantly larger than classical ones. This is a phenomenon often observed in classical-quantum comparisons (see e.g. \cite{Ayadi17}) and may be attributed to quantum tunneling that allows one to explore the momentum space forbidden by classical mechanics.	

In addition, an obvious signature of very strong electronic repulsion in the aligned electron model can be found by analyzing momentum distributions.
In Fig.~\ref{Momenta}(c,f) one can see the pronounced low probability area along the $p_{z1}=p_{z2}$ direction. Releasing electrons with the same momenta would mean having their position close to each other, which is quite unlikely due to the electrons repulsion.

	\section{Conclusions}
	\label{sec:co}

As motivated in the introduction, a study of two electron ionization in strong laser fields at infrared frequencies
is beyond the reach of full quantum simulations. There are several possible reduced-dimensionality models with different
features and different numbers of dimensions. We have compare in our work three most popular choices for such models.
The aligned electrons model (Rochester model) gives unrealistic
electron momentum distributions because the Coulomb repulsion suppresses the experimentally observed dominance of 
equal-momentum events. Here the Eckhardt-Sacha model provides a significant improvement, with the same
number of degrees of freedom and the same numerical complexity. Moreover, it 
performs remarkably well in comparison with the ADK model for the calculation of electron ionization yields.

The Center of Mass model has one degree of freedom more (three instead of two) and necessarily it is more demanding
in terms of the computer time and memory.The ionization yields given by both Eckhardt-Sacha and Center of Mass
models give qualitative trends of similar accuracy. Single electron yields are also in qualitative agreement with 
ADK theory. For momenta distribution the Center of Mass model yields results in closer agreement with experiments,
for Eckardt-Sacha model additional smoothing simulating experimental resolution is needed to remove a pronounced 
interference pattern.

Most of the differences between those two models
can be understood with the application of geometric reasoning. In this sense, the classical trajectory simulations of 
multiple electron ionization provided evidence for the above explanations. 
The calculations have also shown that classical trajectory simulations can reproduce 
the distributions of electron momenta in strong fields. 

In our analysis of the models, we have focused on the two most of the important observables: 
the ionization yield and the final electron momenta distribution.
The case of higher harmonic generation requires further studies and will be pursued in 
a future publication.

	\section{Acknowledgements}
We are grateful to Joe Eberly for friendly remarks at the early stage of this work and to Jesse Mumford for his help with preparing the manuscript. A support by  PL-Grid Infrastructure is acknowledged.
This work was realized under   National Science Center (Poland) project Symfonia  No. 2016/20/W/ST4/00314 (DKE,JPB, ML and JZ) and 2015/19/B/ST2/01028 (AM). 
ML acknowledges also MINECO (National Plan 15
 Grant: FISICATEAMO No. FIS2016-79508-P, SEVERO OCHOA No. SEV-2015-0522,
FOQUS No. FIS2013-46768-P), Fundaci\'o Privada Cellex,  Generalitat de Catalunya (AGAUR
Grant No. 2017 SGR1341 and CERCA/Program), ERC AdG OSYRIS and  EU FETPRO
QUIC.
A.C. thanks to Los Alamos National Laboratory, which is operated by LANS, LLC, for the NNSA of the US DOE under Contract No. DE-AC52-06NA25396. 

\bibliography{reference}

\providecommand{\noopsort}[1]{}\providecommand{\singleletter}[1]{#1}%
\begin{thebibliography}{64}%
\makeatletter
\providecommand \@ifxundefined [1]{%
 \@ifx{#1\undefined}
}%
\providecommand \@ifnum [1]{%
 \ifnum #1\expandafter \@firstoftwo
 \else \expandafter \@secondoftwo
 \fi
}%
\providecommand \@ifx [1]{%
 \ifx #1\expandafter \@firstoftwo
 \else \expandafter \@secondoftwo
 \fi
}%
\providecommand \natexlab [1]{#1}%
\providecommand \enquote  [1]{``#1''}%
\providecommand \bibnamefont  [1]{#1}%
\providecommand \bibfnamefont [1]{#1}%
\providecommand \citenamefont [1]{#1}%
\providecommand \href@noop [0]{\@secondoftwo}%
\providecommand \href [0]{\begingroup \@sanitize@url \@href}%
\providecommand \@href[1]{\@@startlink{#1}\@@href}%
\providecommand \@@href[1]{\endgroup#1\@@endlink}%
\providecommand \@sanitize@url [0]{\catcode `\\12\catcode `\$12\catcode
  `\&12\catcode `\#12\catcode `\^12\catcode `\_12\catcode `\%12\relax}%
\providecommand \@@startlink[1]{}%
\providecommand \@@endlink[0]{}%
\providecommand \url  [0]{\begingroup\@sanitize@url \@url }%
\providecommand \@url [1]{\endgroup\@href {#1}{\urlprefix }}%
\providecommand \urlprefix  [0]{URL }%
\providecommand \Eprint [0]{\href }%
\providecommand \doibase [0]{http://dx.doi.org/}%
\providecommand \selectlanguage [0]{\@gobble}%
\providecommand \bibinfo  [0]{\@secondoftwo}%
\providecommand \bibfield  [0]{\@secondoftwo}%
\providecommand \translation [1]{[#1]}%
\providecommand \BibitemOpen [0]{}%
\providecommand \bibitemStop [0]{}%
\providecommand \bibitemNoStop [0]{.\EOS\space}%
\providecommand \EOS [0]{\spacefactor3000\relax}%
\providecommand \BibitemShut  [1]{\csname bibitem#1\endcsname}%
\let\auto@bib@innerbib\@empty
\bibitem [{\citenamefont {Parker}\ \emph {et~al.}(1998)\citenamefont {Parker},
  \citenamefont {Smyth},\ and\ \citenamefont {Taylor}}]{parker1998intense}%
  \BibitemOpen
  \bibfield  {author} {\bibinfo {author} {\bibfnamefont {J.~S.}\ \bibnamefont
  {Parker}}, \bibinfo {author} {\bibfnamefont {E.~S.}\ \bibnamefont {Smyth}}, \
  and\ \bibinfo {author} {\bibfnamefont {K.~T.}\ \bibnamefont {Taylor}},\
  }\href@noop {} {\bibfield  {journal} {\bibinfo  {journal} {J. Phys. B-At.
  Mol. Opt.}\ }\textbf {\bibinfo {volume} {31}},\ \bibinfo {pages} {L571}
  (\bibinfo {year} {1998})}\BibitemShut {NoStop}%
\bibitem [{\citenamefont {Parker}\ \emph {et~al.}(2000)\citenamefont {Parker},
  \citenamefont {Glass}, \citenamefont {Moore}, \citenamefont {Smyth},
  \citenamefont {Taylor},\ and\ \citenamefont {Burke}}]{parker2000time}%
  \BibitemOpen
  \bibfield  {author} {\bibinfo {author} {\bibfnamefont {J.~S.}\ \bibnamefont
  {Parker}}, \bibinfo {author} {\bibfnamefont {D.}~\bibnamefont {Glass}},
  \bibinfo {author} {\bibfnamefont {L.~R.}\ \bibnamefont {Moore}}, \bibinfo
  {author} {\bibfnamefont {E.~S.}\ \bibnamefont {Smyth}}, \bibinfo {author}
  {\bibfnamefont {K.}~\bibnamefont {Taylor}}, \ and\ \bibinfo {author}
  {\bibfnamefont {P.}~\bibnamefont {Burke}},\ }\href@noop {} {\bibfield
  {journal} {\bibinfo  {journal} {J. Phys. B-At. Mol. Opt.}\ }\textbf {\bibinfo
  {volume} {33}},\ \bibinfo {pages} {L239} (\bibinfo {year}
  {2000})}\BibitemShut {NoStop}%
\bibitem [{\citenamefont {Kulander}\ \emph {et~al.}(1992)\citenamefont
  {Kulander}, \citenamefont {Schafer},\ and\ \citenamefont
  {Krause}}]{kulander1992time}%
  \BibitemOpen
  \bibfield  {author} {\bibinfo {author} {\bibfnamefont {K.~C.}\ \bibnamefont
  {Kulander}}, \bibinfo {author} {\bibfnamefont {K.~J.}\ \bibnamefont
  {Schafer}}, \ and\ \bibinfo {author} {\bibfnamefont {J.~L.}\ \bibnamefont
  {Krause}},\ }in\ \href@noop {} {\emph {\bibinfo {booktitle} {Atoms in intense
  laser fields}}}\ (\bibinfo  {publisher} {Academic Press, Inc., and Harcourt
  Brace Jovanovich},\ \bibinfo {address} {New York},\ \bibinfo {year}
  {1992})\BibitemShut {NoStop}%
\bibitem [{\citenamefont {Feist}\ \emph {et~al.}(2008)\citenamefont {Feist},
  \citenamefont {Nagele}, \citenamefont {Pazourek}, \citenamefont {Persson},
  \citenamefont {Schneider}, \citenamefont {Collins},\ and\ \citenamefont
  {Burgd\"orfer}}]{feist2008nonsequential}%
  \BibitemOpen
  \bibfield  {author} {\bibinfo {author} {\bibfnamefont {J.}~\bibnamefont
  {Feist}}, \bibinfo {author} {\bibfnamefont {S.}~\bibnamefont {Nagele}},
  \bibinfo {author} {\bibfnamefont {R.}~\bibnamefont {Pazourek}}, \bibinfo
  {author} {\bibfnamefont {E.}~\bibnamefont {Persson}}, \bibinfo {author}
  {\bibfnamefont {B.~I.}\ \bibnamefont {Schneider}}, \bibinfo {author}
  {\bibfnamefont {L.~A.}\ \bibnamefont {Collins}}, \ and\ \bibinfo {author}
  {\bibfnamefont {J.}~\bibnamefont {Burgd\"orfer}},\ }\href {\doibase
  10.1103/PhysRevA.77.043420} {\bibfield  {journal} {\bibinfo  {journal} {Phys.
  Rev. A}\ }\textbf {\bibinfo {volume} {77}},\ \bibinfo {pages} {043420}
  (\bibinfo {year} {2008})}\BibitemShut {NoStop}%
\bibitem [{\citenamefont {Djiokap}\ \emph {et~al.}(2017)\citenamefont
  {Djiokap}, \citenamefont {Meremianin}, \citenamefont {Manakov}, \citenamefont
  {Hu}, \citenamefont {Madsen},\ and\ \citenamefont
  {Starace}}]{djiokap2017kinematical}%
  \BibitemOpen
  \bibfield  {author} {\bibinfo {author} {\bibfnamefont {J.~M.~N.}\
  \bibnamefont {Djiokap}}, \bibinfo {author} {\bibfnamefont {A.~V.}\
  \bibnamefont {Meremianin}}, \bibinfo {author} {\bibfnamefont {N.~L.}\
  \bibnamefont {Manakov}}, \bibinfo {author} {\bibfnamefont {S.~X.}\
  \bibnamefont {Hu}}, \bibinfo {author} {\bibfnamefont {L.~B.}\ \bibnamefont
  {Madsen}}, \ and\ \bibinfo {author} {\bibfnamefont {A.~F.}\ \bibnamefont
  {Starace}},\ }\href {\doibase 10.1103/PhysRevA.96.013405} {\bibfield
  {journal} {\bibinfo  {journal} {Phys. Rev. A}\ }\textbf {\bibinfo {volume}
  {96}},\ \bibinfo {pages} {013405} (\bibinfo {year} {2017})}\BibitemShut
  {NoStop}%
\bibitem [{\citenamefont {Becker}\ \emph {et~al.}(2012)\citenamefont {Becker},
  \citenamefont {Liu}, \citenamefont {Ho},\ and\ \citenamefont
  {Eberly}}]{becker2012theories}%
  \BibitemOpen
  \bibfield  {author} {\bibinfo {author} {\bibfnamefont {W.}~\bibnamefont
  {Becker}}, \bibinfo {author} {\bibfnamefont {X.}~\bibnamefont {Liu}},
  \bibinfo {author} {\bibfnamefont {P.~J.}\ \bibnamefont {Ho}}, \ and\ \bibinfo
  {author} {\bibfnamefont {J.~H.}\ \bibnamefont {Eberly}},\ }\href@noop {}
  {\bibfield  {journal} {\bibinfo  {journal} {Rev. Mod. Phys.}\ }\textbf
  {\bibinfo {volume} {84}},\ \bibinfo {pages} {1011} (\bibinfo {year}
  {2012})}\BibitemShut {NoStop}%
\bibitem [{\citenamefont {Walker}\ \emph {et~al.}(1994)\citenamefont {Walker},
  \citenamefont {Sheehy}, \citenamefont {DiMauro}, \citenamefont {Agostini},
  \citenamefont {Schafer},\ and\ \citenamefont {Kulander}}]{Walker94}%
  \BibitemOpen
  \bibfield  {author} {\bibinfo {author} {\bibfnamefont {B.}~\bibnamefont
  {Walker}}, \bibinfo {author} {\bibfnamefont {B.}~\bibnamefont {Sheehy}},
  \bibinfo {author} {\bibfnamefont {L.~F.}\ \bibnamefont {DiMauro}}, \bibinfo
  {author} {\bibfnamefont {P.}~\bibnamefont {Agostini}}, \bibinfo {author}
  {\bibfnamefont {K.~J.}\ \bibnamefont {Schafer}}, \ and\ \bibinfo {author}
  {\bibfnamefont {K.~C.}\ \bibnamefont {Kulander}},\ }\href {\doibase
  10.1103/PhysRevLett.73.1227} {\bibfield  {journal} {\bibinfo  {journal}
  {Phys. Rev. Lett.}\ }\textbf {\bibinfo {volume} {73}},\ \bibinfo {pages}
  {1227} (\bibinfo {year} {1994})}\BibitemShut {NoStop}%
\bibitem [{\citenamefont {Fittinghoff}\ \emph {et~al.}(1992)\citenamefont
  {Fittinghoff}, \citenamefont {Bolton}, \citenamefont {Chang},\ and\
  \citenamefont {Kulander}}]{fittinghoff1992observation}%
  \BibitemOpen
  \bibfield  {author} {\bibinfo {author} {\bibfnamefont {D.~N.}\ \bibnamefont
  {Fittinghoff}}, \bibinfo {author} {\bibfnamefont {P.~R.}\ \bibnamefont
  {Bolton}}, \bibinfo {author} {\bibfnamefont {B.}~\bibnamefont {Chang}}, \
  and\ \bibinfo {author} {\bibfnamefont {K.~C.}\ \bibnamefont {Kulander}},\
  }\href@noop {} {\bibfield  {journal} {\bibinfo  {journal} {Phys. Rev. Lett.}\
  }\textbf {\bibinfo {volume} {69}},\ \bibinfo {pages} {2642} (\bibinfo {year}
  {1992})}\BibitemShut {NoStop}%
\bibitem [{\citenamefont {Lein}\ \emph {et~al.}(2000)\citenamefont {Lein},
  \citenamefont {Gross},\ and\ \citenamefont {Engel}}]{lein2000intense}%
  \BibitemOpen
  \bibfield  {author} {\bibinfo {author} {\bibfnamefont {M.}~\bibnamefont
  {Lein}}, \bibinfo {author} {\bibfnamefont {E.~K.~U.}\ \bibnamefont {Gross}},
  \ and\ \bibinfo {author} {\bibfnamefont {V.}~\bibnamefont {Engel}},\ }\href
  {\doibase 10.1103/PhysRevLett.85.4707} {\bibfield  {journal} {\bibinfo
  {journal} {Phys. Rev. Lett.}\ }\textbf {\bibinfo {volume} {85}},\ \bibinfo
  {pages} {4707} (\bibinfo {year} {2000})}\BibitemShut {NoStop}%
\bibitem [{\citenamefont {Ruiz}\ \emph {et~al.}(2003)\citenamefont {Ruiz},
  \citenamefont {Plaja}, \citenamefont {V\'azquez~de Aldana},\ and\
  \citenamefont {Roso}}]{ruizPRA03}%
  \BibitemOpen
  \bibfield  {author} {\bibinfo {author} {\bibfnamefont {C.}~\bibnamefont
  {Ruiz}}, \bibinfo {author} {\bibfnamefont {L.}~\bibnamefont {Plaja}},
  \bibinfo {author} {\bibfnamefont {J.~R.}\ \bibnamefont {V\'azquez~de
  Aldana}}, \ and\ \bibinfo {author} {\bibfnamefont {L.}~\bibnamefont {Roso}},\
  }\href@noop {} {\bibfield  {journal} {\bibinfo  {journal} {Phys. Rev. A}\
  }\textbf {\bibinfo {volume} {68}},\ \bibinfo {pages} {023409} (\bibinfo
  {year} {2003})}\BibitemShut {NoStop}%
\bibitem [{\citenamefont {Prauzner-Bechcicki}\ \emph
  {et~al.}(2007)\citenamefont {Prauzner-Bechcicki}, \citenamefont {Sacha},
  \citenamefont {Eckhardt},\ and\ \citenamefont
  {Zakrzewski}}]{prauzner2007time}%
  \BibitemOpen
  \bibfield  {author} {\bibinfo {author} {\bibfnamefont {J.~S.}\ \bibnamefont
  {Prauzner-Bechcicki}}, \bibinfo {author} {\bibfnamefont {K.}~\bibnamefont
  {Sacha}}, \bibinfo {author} {\bibfnamefont {B.}~\bibnamefont {Eckhardt}}, \
  and\ \bibinfo {author} {\bibfnamefont {J.}~\bibnamefont {Zakrzewski}},\
  }\href@noop {} {\bibfield  {journal} {\bibinfo  {journal} {Phys. Rev. Lett.}\
  }\textbf {\bibinfo {volume} {98}},\ \bibinfo {pages} {203002} (\bibinfo
  {year} {2007})}\BibitemShut {NoStop}%
\bibitem [{\citenamefont {Prauzner-Bechcicki}\ \emph
  {et~al.}(2008)\citenamefont {Prauzner-Bechcicki}, \citenamefont {Sacha},
  \citenamefont {Eckhardt},\ and\ \citenamefont
  {Zakrzewski}}]{prauzner2008quantum}%
  \BibitemOpen
  \bibfield  {author} {\bibinfo {author} {\bibfnamefont {J.~S.}\ \bibnamefont
  {Prauzner-Bechcicki}}, \bibinfo {author} {\bibfnamefont {K.}~\bibnamefont
  {Sacha}}, \bibinfo {author} {\bibfnamefont {B.}~\bibnamefont {Eckhardt}}, \
  and\ \bibinfo {author} {\bibfnamefont {J.}~\bibnamefont {Zakrzewski}},\
  }\href@noop {} {\bibfield  {journal} {\bibinfo  {journal} {Phys. Rev. A}\
  }\textbf {\bibinfo {volume} {78}},\ \bibinfo {pages} {013419} (\bibinfo
  {year} {2008})}\BibitemShut {NoStop}%
\bibitem [{\citenamefont {Panfili}\ \emph
  {et~al.}(2001{\natexlab{a}})\citenamefont {Panfili}, \citenamefont {Eberly},\
  and\ \citenamefont {Haan}}]{panfili2001comparing}%
  \BibitemOpen
  \bibfield  {author} {\bibinfo {author} {\bibfnamefont {R.}~\bibnamefont
  {Panfili}}, \bibinfo {author} {\bibfnamefont {J.~H.}\ \bibnamefont {Eberly}},
  \ and\ \bibinfo {author} {\bibfnamefont {S.~L.}\ \bibnamefont {Haan}},\
  }\href@noop {} {\bibfield  {journal} {\bibinfo  {journal} {Opt. Express}\
  }\textbf {\bibinfo {volume} {8}},\ \bibinfo {pages} {431} (\bibinfo {year}
  {2001}{\natexlab{a}})}\BibitemShut {NoStop}%
\bibitem [{\citenamefont {Panfili}\ \emph
  {et~al.}(2002{\natexlab{a}})\citenamefont {Panfili}, \citenamefont {Haan},\
  and\ \citenamefont {Eberly}}]{panfili2002slow}%
  \BibitemOpen
  \bibfield  {author} {\bibinfo {author} {\bibfnamefont {R.}~\bibnamefont
  {Panfili}}, \bibinfo {author} {\bibfnamefont {S.~L.}\ \bibnamefont {Haan}}, \
  and\ \bibinfo {author} {\bibfnamefont {J.~H.}\ \bibnamefont {Eberly}},\
  }\href@noop {} {\bibfield  {journal} {\bibinfo  {journal} {Phys. Rev. Lett.}\
  }\textbf {\bibinfo {volume} {89}},\ \bibinfo {pages} {113001} (\bibinfo
  {year} {2002}{\natexlab{a}})}\BibitemShut {NoStop}%
\bibitem [{\citenamefont {Weber}\ \emph {et~al.}(2000)\citenamefont {Weber},
  \citenamefont {Giessen}, \citenamefont {Weckenbrock}, \citenamefont
  {Urbasch}, \citenamefont {Staudte}, \citenamefont {Spielberger},
  \citenamefont {Jagutzki}, \citenamefont {Mergel}, \citenamefont {Vollmer},\
  and\ \citenamefont {D{\"o}rner}}]{weber2000correlated}%
  \BibitemOpen
  \bibfield  {author} {\bibinfo {author} {\bibfnamefont {T.}~\bibnamefont
  {Weber}}, \bibinfo {author} {\bibfnamefont {H.}~\bibnamefont {Giessen}},
  \bibinfo {author} {\bibfnamefont {M.}~\bibnamefont {Weckenbrock}}, \bibinfo
  {author} {\bibfnamefont {G.}~\bibnamefont {Urbasch}}, \bibinfo {author}
  {\bibfnamefont {A.}~\bibnamefont {Staudte}}, \bibinfo {author} {\bibfnamefont
  {L.}~\bibnamefont {Spielberger}}, \bibinfo {author} {\bibfnamefont
  {O.}~\bibnamefont {Jagutzki}}, \bibinfo {author} {\bibfnamefont
  {V.}~\bibnamefont {Mergel}}, \bibinfo {author} {\bibfnamefont
  {M.}~\bibnamefont {Vollmer}}, \ and\ \bibinfo {author} {\bibfnamefont
  {R.}~\bibnamefont {D{\"o}rner}},\ }\href@noop {} {\bibfield  {journal}
  {\bibinfo  {journal} {Nature}\ }\textbf {\bibinfo {volume} {405}},\ \bibinfo
  {pages} {658} (\bibinfo {year} {2000})}\BibitemShut {NoStop}%
\bibitem [{\citenamefont {Rudenko}\ \emph {et~al.}(2007)\citenamefont
  {Rudenko}, \citenamefont {de~Jesus}, \citenamefont {Ergler}, \citenamefont
  {Zrost}, \citenamefont {Feuerstein}, \citenamefont {Schr\"oter},
  \citenamefont {Moshammer},\ and\ \citenamefont
  {Ullrich}}]{rudenko2007correlated}%
  \BibitemOpen
  \bibfield  {author} {\bibinfo {author} {\bibfnamefont {A.}~\bibnamefont
  {Rudenko}}, \bibinfo {author} {\bibfnamefont {V.~L.~B.}\ \bibnamefont
  {de~Jesus}}, \bibinfo {author} {\bibfnamefont {T.}~\bibnamefont {Ergler}},
  \bibinfo {author} {\bibfnamefont {K.}~\bibnamefont {Zrost}}, \bibinfo
  {author} {\bibfnamefont {B.}~\bibnamefont {Feuerstein}}, \bibinfo {author}
  {\bibfnamefont {C.~D.}\ \bibnamefont {Schr\"oter}}, \bibinfo {author}
  {\bibfnamefont {R.}~\bibnamefont {Moshammer}}, \ and\ \bibinfo {author}
  {\bibfnamefont {J.}~\bibnamefont {Ullrich}},\ }\href {\doibase
  10.1103/PhysRevLett.99.263003} {\bibfield  {journal} {\bibinfo  {journal}
  {Phys. Rev. Lett.}\ }\textbf {\bibinfo {volume} {99}},\ \bibinfo {pages}
  {263003} (\bibinfo {year} {2007})}\BibitemShut {NoStop}%
\bibitem [{\citenamefont {K{\"u}bel}\ \emph {et~al.}(2014)\citenamefont
  {K{\"u}bel}, \citenamefont {Betsch}, \citenamefont {Kling}, \citenamefont
  {Alnaser}, \citenamefont {Schmidt}, \citenamefont {Kleineberg}, \citenamefont
  {Deng}, \citenamefont {Ben-Itzhak}, \citenamefont {Paulus}, \citenamefont
  {Pfeifer} \emph {et~al.}}]{kubel2014non}%
  \BibitemOpen
  \bibfield  {author} {\bibinfo {author} {\bibfnamefont {M.}~\bibnamefont
  {K{\"u}bel}}, \bibinfo {author} {\bibfnamefont {K.}~\bibnamefont {Betsch}},
  \bibinfo {author} {\bibfnamefont {N.~G.}\ \bibnamefont {Kling}}, \bibinfo
  {author} {\bibfnamefont {A.}~\bibnamefont {Alnaser}}, \bibinfo {author}
  {\bibfnamefont {J.}~\bibnamefont {Schmidt}}, \bibinfo {author} {\bibfnamefont
  {U.}~\bibnamefont {Kleineberg}}, \bibinfo {author} {\bibfnamefont
  {Y.}~\bibnamefont {Deng}}, \bibinfo {author} {\bibfnamefont {I.}~\bibnamefont
  {Ben-Itzhak}}, \bibinfo {author} {\bibfnamefont {G.}~\bibnamefont {Paulus}},
  \bibinfo {author} {\bibfnamefont {T.}~\bibnamefont {Pfeifer}},  \emph
  {et~al.},\ }\href@noop {} {\bibfield  {journal} {\bibinfo  {journal} {New J.
  Phys.}\ }\textbf {\bibinfo {volume} {16}},\ \bibinfo {pages} {033008}
  (\bibinfo {year} {2014})}\BibitemShut {NoStop}%
\bibitem [{\citenamefont {Camus}\ \emph {et~al.}(2012)\citenamefont {Camus},
  \citenamefont {Fischer}, \citenamefont {Kremer}, \citenamefont {Sharma},
  \citenamefont {Rudenko}, \citenamefont {Bergues}, \citenamefont {K{\"u}bel},
  \citenamefont {Johnson}, \citenamefont {Kling}, \citenamefont {Pfeifer} \emph
  {et~al.}}]{camus2012attosecond}%
  \BibitemOpen
  \bibfield  {author} {\bibinfo {author} {\bibfnamefont {N.}~\bibnamefont
  {Camus}}, \bibinfo {author} {\bibfnamefont {B.}~\bibnamefont {Fischer}},
  \bibinfo {author} {\bibfnamefont {M.}~\bibnamefont {Kremer}}, \bibinfo
  {author} {\bibfnamefont {V.}~\bibnamefont {Sharma}}, \bibinfo {author}
  {\bibfnamefont {A.}~\bibnamefont {Rudenko}}, \bibinfo {author} {\bibfnamefont
  {B.}~\bibnamefont {Bergues}}, \bibinfo {author} {\bibfnamefont
  {M.}~\bibnamefont {K{\"u}bel}}, \bibinfo {author} {\bibfnamefont {N.~G.}\
  \bibnamefont {Johnson}}, \bibinfo {author} {\bibfnamefont {M.~F.}\
  \bibnamefont {Kling}}, \bibinfo {author} {\bibfnamefont {T.}~\bibnamefont
  {Pfeifer}},  \emph {et~al.},\ }\href@noop {} {\bibfield  {journal} {\bibinfo
  {journal} {Phys. Rev. Lett.}\ }\textbf {\bibinfo {volume} {108}},\ \bibinfo
  {pages} {073003} (\bibinfo {year} {2012})}\BibitemShut {NoStop}%
\bibitem [{\citenamefont {Staudte}\ \emph {et~al.}(2007)\citenamefont
  {Staudte}, \citenamefont {Ruiz}, \citenamefont {Sch{\"o}ffler}, \citenamefont
  {Sch{\"o}ssler}, \citenamefont {Zeidler}, \citenamefont {Weber},
  \citenamefont {Meckel}, \citenamefont {Villeneuve}, \citenamefont {Corkum},
  \citenamefont {Becker} \emph {et~al.}}]{staudte2007binary}%
  \BibitemOpen
  \bibfield  {author} {\bibinfo {author} {\bibfnamefont {A.}~\bibnamefont
  {Staudte}}, \bibinfo {author} {\bibfnamefont {C.}~\bibnamefont {Ruiz}},
  \bibinfo {author} {\bibfnamefont {M.}~\bibnamefont {Sch{\"o}ffler}}, \bibinfo
  {author} {\bibfnamefont {S.}~\bibnamefont {Sch{\"o}ssler}}, \bibinfo {author}
  {\bibfnamefont {D.}~\bibnamefont {Zeidler}}, \bibinfo {author} {\bibfnamefont
  {T.}~\bibnamefont {Weber}}, \bibinfo {author} {\bibfnamefont
  {M.}~\bibnamefont {Meckel}}, \bibinfo {author} {\bibfnamefont
  {D.}~\bibnamefont {Villeneuve}}, \bibinfo {author} {\bibfnamefont
  {P.}~\bibnamefont {Corkum}}, \bibinfo {author} {\bibfnamefont
  {A.}~\bibnamefont {Becker}},  \emph {et~al.},\ }\href@noop {} {\bibfield
  {journal} {\bibinfo  {journal} {Phys. Rev. Lett.}\ }\textbf {\bibinfo
  {volume} {99}},\ \bibinfo {pages} {263002} (\bibinfo {year}
  {2007})}\BibitemShut {NoStop}%
\bibitem [{\citenamefont {Ye}\ \emph {et~al.}(2008)\citenamefont {Ye},
  \citenamefont {Liu},\ and\ \citenamefont {Liu}}]{ye2008classical}%
  \BibitemOpen
  \bibfield  {author} {\bibinfo {author} {\bibfnamefont {D.~F.}\ \bibnamefont
  {Ye}}, \bibinfo {author} {\bibfnamefont {X.}~\bibnamefont {Liu}}, \ and\
  \bibinfo {author} {\bibfnamefont {J.}~\bibnamefont {Liu}},\ }\href {\doibase
  10.1103/PhysRevLett.101.233003} {\bibfield  {journal} {\bibinfo  {journal}
  {Phys. Rev. Lett.}\ }\textbf {\bibinfo {volume} {101}},\ \bibinfo {pages}
  {233003} (\bibinfo {year} {2008})}\BibitemShut {NoStop}%
\bibitem [{\citenamefont {Koval}\ \emph {et~al.}(2007)\citenamefont {Koval},
  \citenamefont {Wilken}, \citenamefont {Bauer},\ and\ \citenamefont
  {Keitel}}]{koval2007nonsequential}%
  \BibitemOpen
  \bibfield  {author} {\bibinfo {author} {\bibfnamefont {P.}~\bibnamefont
  {Koval}}, \bibinfo {author} {\bibfnamefont {F.}~\bibnamefont {Wilken}},
  \bibinfo {author} {\bibfnamefont {D.}~\bibnamefont {Bauer}}, \ and\ \bibinfo
  {author} {\bibfnamefont {C.~H.}\ \bibnamefont {Keitel}},\ }\href {\doibase
  10.1103/PhysRevLett.98.043904} {\bibfield  {journal} {\bibinfo  {journal}
  {Phys. Rev. Lett.}\ }\textbf {\bibinfo {volume} {98}},\ \bibinfo {pages}
  {043904} (\bibinfo {year} {2007})}\BibitemShut {NoStop}%
\bibitem [{\citenamefont {Bauer}(1997)}]{bauer1997two}%
  \BibitemOpen
  \bibfield  {author} {\bibinfo {author} {\bibfnamefont {D.}~\bibnamefont
  {Bauer}},\ }\href@noop {} {\bibfield  {journal} {\bibinfo  {journal} {Phys.
  Rev. A}\ }\textbf {\bibinfo {volume} {56}},\ \bibinfo {pages} {3028}
  (\bibinfo {year} {1997})}\BibitemShut {NoStop}%
\bibitem [{\citenamefont {Chen}\ \emph
  {et~al.}(2018{\natexlab{a}})\citenamefont {Chen}, \citenamefont {Zhou},
  \citenamefont {Li}, \citenamefont {Li}, \citenamefont {Lan},\ and\
  \citenamefont {Lu}}]{chen2018rabi}%
  \BibitemOpen
  \bibfield  {author} {\bibinfo {author} {\bibfnamefont {Y.}~\bibnamefont
  {Chen}}, \bibinfo {author} {\bibfnamefont {Y.}~\bibnamefont {Zhou}}, \bibinfo
  {author} {\bibfnamefont {Y.}~\bibnamefont {Li}}, \bibinfo {author}
  {\bibfnamefont {M.}~\bibnamefont {Li}}, \bibinfo {author} {\bibfnamefont
  {P.}~\bibnamefont {Lan}}, \ and\ \bibinfo {author} {\bibfnamefont
  {P.}~\bibnamefont {Lu}},\ }\href@noop {} {\bibfield  {journal} {\bibinfo
  {journal} {Phys. Rev. A}\ }\textbf {\bibinfo {volume} {97}},\ \bibinfo
  {pages} {013428} (\bibinfo {year} {2018}{\natexlab{a}})}\BibitemShut
  {NoStop}%
\bibitem [{\citenamefont {Sacha}\ and\ \citenamefont
  {Eckhardt}(2001)}]{sacha2001pathways}%
  \BibitemOpen
  \bibfield  {author} {\bibinfo {author} {\bibfnamefont {K.}~\bibnamefont
  {Sacha}}\ and\ \bibinfo {author} {\bibfnamefont {B.}~\bibnamefont
  {Eckhardt}},\ }\href@noop {} {\bibfield  {journal} {\bibinfo  {journal}
  {Phys. Rev. A}\ }\textbf {\bibinfo {volume} {63}},\ \bibinfo {pages} {043414}
  (\bibinfo {year} {2001})}\BibitemShut {NoStop}%
\bibitem [{\citenamefont {Eckhardt}\ \emph {et~al.}(2010)\citenamefont
  {Eckhardt}, \citenamefont {Prauzner-Bechcicki}, \citenamefont {Sacha},\ and\
  \citenamefont {Zakrzewski}}]{eckhardt2010phase}%
  \BibitemOpen
  \bibfield  {author} {\bibinfo {author} {\bibfnamefont {B.}~\bibnamefont
  {Eckhardt}}, \bibinfo {author} {\bibfnamefont {J.~S.}\ \bibnamefont
  {Prauzner-Bechcicki}}, \bibinfo {author} {\bibfnamefont {K.}~\bibnamefont
  {Sacha}}, \ and\ \bibinfo {author} {\bibfnamefont {J.}~\bibnamefont
  {Zakrzewski}},\ }\href@noop {} {\bibfield  {journal} {\bibinfo  {journal}
  {Chem. Phys.}\ }\textbf {\bibinfo {volume} {370}},\ \bibinfo {pages} {168}
  (\bibinfo {year} {2010})}\BibitemShut {NoStop}%
\bibitem [{\citenamefont {Ruiz}\ \emph {et~al.}(2006)\citenamefont {Ruiz},
  \citenamefont {Plaja}, \citenamefont {Roso},\ and\ \citenamefont
  {Becker}}]{ruiz2006ab}%
  \BibitemOpen
  \bibfield  {author} {\bibinfo {author} {\bibfnamefont {C.}~\bibnamefont
  {Ruiz}}, \bibinfo {author} {\bibfnamefont {L.}~\bibnamefont {Plaja}},
  \bibinfo {author} {\bibfnamefont {L.}~\bibnamefont {Roso}}, \ and\ \bibinfo
  {author} {\bibfnamefont {A.}~\bibnamefont {Becker}},\ }\href@noop {}
  {\bibfield  {journal} {\bibinfo  {journal} {Phys. Rev. Lett.}\ }\textbf
  {\bibinfo {volume} {96}},\ \bibinfo {pages} {053001} (\bibinfo {year}
  {2006})}\BibitemShut {NoStop}%
\bibitem [{\citenamefont {Leopold}\ and\ \citenamefont
  {Percival}(1979)}]{Leopold79}%
  \BibitemOpen
  \bibfield  {author} {\bibinfo {author} {\bibfnamefont {J.}~\bibnamefont
  {Leopold}}\ and\ \bibinfo {author} {\bibfnamefont {I.~C.}\ \bibnamefont
  {Percival}},\ }\href@noop {} {\bibfield  {journal} {\bibinfo  {journal} {J.
  Phys. B-At. Mol. Opt.}\ }\textbf {\bibinfo {volume} {12}},\ \bibinfo {pages}
  {709} (\bibinfo {year} {1979})}\BibitemShut {NoStop}%
\bibitem [{\citenamefont {Grochmalicki}\ \emph {et~al.}(1991)\citenamefont
  {Grochmalicki}, \citenamefont {Lewenstein},\ and\ \citenamefont
  {Rza\ifmmode~\mbox{\c{}}\else
  \c{}\fi{}ewski}}]{grochmalicki1991stabilization}%
  \BibitemOpen
  \bibfield  {author} {\bibinfo {author} {\bibfnamefont {J.}~\bibnamefont
  {Grochmalicki}}, \bibinfo {author} {\bibfnamefont {M.}~\bibnamefont
  {Lewenstein}}, \ and\ \bibinfo {author} {\bibfnamefont {K.}~\bibnamefont
  {Rza\ifmmode~\mbox{\c{}}\else \c{}\fi{}ewski}},\ }\href {\doibase
  10.1103/PhysRevLett.66.1038} {\bibfield  {journal} {\bibinfo  {journal}
  {Phys. Rev. Lett.}\ }\textbf {\bibinfo {volume} {66}},\ \bibinfo {pages}
  {1038} (\bibinfo {year} {1991})}\BibitemShut {NoStop}%
\bibitem [{\citenamefont {Gajda}\ \emph {et~al.}(1992)\citenamefont {Gajda},
  \citenamefont {Grochmalicki}, \citenamefont {Lewenstein},\ and\ \citenamefont
  {Rza\ifmmode \mbox{\c{}}\else \c{}\fi{}\ifmmode~\dot{z}\else
  \.{z}\fi{}ewski}}]{Gajda92}%
  \BibitemOpen
  \bibfield  {author} {\bibinfo {author} {\bibfnamefont {M.}~\bibnamefont
  {Gajda}}, \bibinfo {author} {\bibfnamefont {J.}~\bibnamefont {Grochmalicki}},
  \bibinfo {author} {\bibfnamefont {M.}~\bibnamefont {Lewenstein}}, \ and\
  \bibinfo {author} {\bibfnamefont {K.}~\bibnamefont {Rza\ifmmode
  \mbox{\c{}}\else \c{}\fi{}\ifmmode~\dot{z}\else \.{z}\fi{}ewski}},\ }\href
  {\doibase 10.1103/PhysRevA.46.1638} {\bibfield  {journal} {\bibinfo
  {journal} {Phys. Rev. A}\ }\textbf {\bibinfo {volume} {46}},\ \bibinfo
  {pages} {1638} (\bibinfo {year} {1992})}\BibitemShut {NoStop}%
\bibitem [{\citenamefont {Rza\ifmmode \mbox{\c{}}\else
  \c{}\fi{}\ifmmode~\dot{z}\else \.{z}\fi{}ewski}\ \emph
  {et~al.}(1994)\citenamefont {Rza\ifmmode \mbox{\c{}}\else
  \c{}\fi{}\ifmmode~\dot{z}\else \.{z}\fi{}ewski}, \citenamefont {Lewenstein},\
  and\ \citenamefont {Sali\`eres}}]{Rzazewski94}%
  \BibitemOpen
  \bibfield  {author} {\bibinfo {author} {\bibfnamefont {K.}~\bibnamefont
  {Rza\ifmmode \mbox{\c{}}\else \c{}\fi{}\ifmmode~\dot{z}\else
  \.{z}\fi{}ewski}}, \bibinfo {author} {\bibfnamefont {M.}~\bibnamefont
  {Lewenstein}}, \ and\ \bibinfo {author} {\bibfnamefont {P.}~\bibnamefont
  {Sali\`eres}},\ }\href {\doibase 10.1103/PhysRevA.49.1196} {\bibfield
  {journal} {\bibinfo  {journal} {Phys. Rev. A}\ }\textbf {\bibinfo {volume}
  {49}},\ \bibinfo {pages} {1196} (\bibinfo {year} {1994})}\BibitemShut
  {NoStop}%
\bibitem [{\citenamefont {W\'ojcik}\ \emph {et~al.}(1995)\citenamefont
  {W\'ojcik}, \citenamefont {Zakrzewski},\ and\ \citenamefont {Rza\ifmmode
  \mbox{\c{}}\else \c{}\fi{}\ifmmode~\dot{z}\else \.{z}\fi{}ewski}}]{Wojcik95}%
  \BibitemOpen
  \bibfield  {author} {\bibinfo {author} {\bibfnamefont {M.}~\bibnamefont
  {W\'ojcik}}, \bibinfo {author} {\bibfnamefont {J.}~\bibnamefont
  {Zakrzewski}}, \ and\ \bibinfo {author} {\bibfnamefont {K.}~\bibnamefont
  {Rza\ifmmode \mbox{\c{}}\else \c{}\fi{}\ifmmode~\dot{z}\else
  \.{z}\fi{}ewski}},\ }\href {\doibase 10.1103/PhysRevA.52.R2523} {\bibfield
  {journal} {\bibinfo  {journal} {Phys. Rev. A}\ }\textbf {\bibinfo {volume}
  {52}},\ \bibinfo {pages} {R2523} (\bibinfo {year} {1995})}\BibitemShut
  {NoStop}%
\bibitem [{\citenamefont {Ho}\ and\ \citenamefont
  {Eberly}(2007)}]{ho2007argon}%
  \BibitemOpen
  \bibfield  {author} {\bibinfo {author} {\bibfnamefont {P.~J.}\ \bibnamefont
  {Ho}}\ and\ \bibinfo {author} {\bibfnamefont {J.}~\bibnamefont {Eberly}},\
  }\href@noop {} {\bibfield  {journal} {\bibinfo  {journal} {Opt. Express}\
  }\textbf {\bibinfo {volume} {15}},\ \bibinfo {pages} {1845} (\bibinfo {year}
  {2007})}\BibitemShut {NoStop}%
\bibitem [{\citenamefont {Chen}\ \emph {et~al.}(2000)\citenamefont {Chen},
  \citenamefont {Liu}, \citenamefont {Fu},\ and\ \citenamefont
  {Zheng}}]{chen2000interpretation}%
  \BibitemOpen
  \bibfield  {author} {\bibinfo {author} {\bibfnamefont {J.}~\bibnamefont
  {Chen}}, \bibinfo {author} {\bibfnamefont {J.}~\bibnamefont {Liu}}, \bibinfo
  {author} {\bibfnamefont {L.~B.}\ \bibnamefont {Fu}}, \ and\ \bibinfo {author}
  {\bibfnamefont {W.~M.}\ \bibnamefont {Zheng}},\ }\href {\doibase
  10.1103/PhysRevA.63.011404} {\bibfield  {journal} {\bibinfo  {journal} {Phys.
  Rev. A}\ }\textbf {\bibinfo {volume} {63}},\ \bibinfo {pages} {011404}
  (\bibinfo {year} {2000})}\BibitemShut {NoStop}%
\bibitem [{\citenamefont {Ho}\ \emph {et~al.}(2005)\citenamefont {Ho},
  \citenamefont {Panfili}, \citenamefont {Haan},\ and\ \citenamefont
  {Eberly}}]{ho2005nonsequential}%
  \BibitemOpen
  \bibfield  {author} {\bibinfo {author} {\bibfnamefont {P.~J.}\ \bibnamefont
  {Ho}}, \bibinfo {author} {\bibfnamefont {R.}~\bibnamefont {Panfili}},
  \bibinfo {author} {\bibfnamefont {S.~L.}\ \bibnamefont {Haan}}, \ and\
  \bibinfo {author} {\bibfnamefont {J.~H.}\ \bibnamefont {Eberly}},\ }\href
  {\doibase 10.1103/PhysRevLett.94.093002} {\bibfield  {journal} {\bibinfo
  {journal} {Phys. Rev. Lett.}\ }\textbf {\bibinfo {volume} {94}},\ \bibinfo
  {pages} {093002} (\bibinfo {year} {2005})}\BibitemShut {NoStop}%
\bibitem [{\citenamefont {Mauger}\ \emph
  {et~al.}(2009{\natexlab{a}})\citenamefont {Mauger}, \citenamefont {Chandre},\
  and\ \citenamefont {Uzer}}]{mauger2009strong}%
  \BibitemOpen
  \bibfield  {author} {\bibinfo {author} {\bibfnamefont {F.}~\bibnamefont
  {Mauger}}, \bibinfo {author} {\bibfnamefont {C.}~\bibnamefont {Chandre}}, \
  and\ \bibinfo {author} {\bibfnamefont {T.}~\bibnamefont {Uzer}},\ }\href@noop
  {} {\bibfield  {journal} {\bibinfo  {journal} {Phys. Rev. Lett.}\ }\textbf
  {\bibinfo {volume} {102}},\ \bibinfo {pages} {173002} (\bibinfo {year}
  {2009}{\natexlab{a}})}\BibitemShut {NoStop}%
\bibitem [{\citenamefont {Berman}\ \emph {et~al.}(2018)\citenamefont {Berman},
  \citenamefont {Chandre}, \citenamefont {Perin},\ and\ \citenamefont
  {Uzer}}]{berman2018coherent}%
  \BibitemOpen
  \bibfield  {author} {\bibinfo {author} {\bibfnamefont {S.}~\bibnamefont
  {Berman}}, \bibinfo {author} {\bibfnamefont {C.}~\bibnamefont {Chandre}},
  \bibinfo {author} {\bibfnamefont {M.}~\bibnamefont {Perin}}, \ and\ \bibinfo
  {author} {\bibfnamefont {T.}~\bibnamefont {Uzer}},\ }\href@noop {} {\bibfield
   {journal} {\bibinfo  {journal} {arXiv preprint arXiv:1801.09727}\ }
  (\bibinfo {year} {2018})}\BibitemShut {NoStop}%
\bibitem [{\citenamefont {Ammosov}\ \emph {et~al.}(1986)\citenamefont
  {Ammosov}, \citenamefont {Delone},\ and\ \citenamefont
  {Krainov}}]{ammosov1986tunnel}%
  \BibitemOpen
  \bibfield  {author} {\bibinfo {author} {\bibfnamefont {M.}~\bibnamefont
  {Ammosov}}, \bibinfo {author} {\bibfnamefont {N.}~\bibnamefont {Delone}}, \
  and\ \bibinfo {author} {\bibnamefont {Krainov}},\ }\href@noop {} {\bibfield
  {journal} {\bibinfo  {journal} {Sov. Phys. JETP}\ }\textbf {\bibinfo {volume}
  {64}},\ \bibinfo {pages} {1191} (\bibinfo {year} {1986})}\BibitemShut
  {NoStop}%
\bibitem [{\citenamefont {Kondo}\ \emph {et~al.}(1993)\citenamefont {Kondo},
  \citenamefont {Sagisaka}, \citenamefont {Tamida}, \citenamefont {Nabekawa},\
  and\ \citenamefont {Watanabe}}]{kondo1993wavelength}%
  \BibitemOpen
  \bibfield  {author} {\bibinfo {author} {\bibfnamefont {K.}~\bibnamefont
  {Kondo}}, \bibinfo {author} {\bibfnamefont {A.}~\bibnamefont {Sagisaka}},
  \bibinfo {author} {\bibfnamefont {T.}~\bibnamefont {Tamida}}, \bibinfo
  {author} {\bibfnamefont {Y.}~\bibnamefont {Nabekawa}}, \ and\ \bibinfo
  {author} {\bibfnamefont {S.}~\bibnamefont {Watanabe}},\ }\href@noop {}
  {\bibfield  {journal} {\bibinfo  {journal} {Phys. Rev. A}\ }\textbf {\bibinfo
  {volume} {48}},\ \bibinfo {pages} {R2531} (\bibinfo {year}
  {1993})}\BibitemShut {NoStop}%
\bibitem [{\citenamefont {Larochelle}\ \emph {et~al.}(1998)\citenamefont
  {Larochelle}, \citenamefont {Talebpour},\ and\ \citenamefont
  {Chin}}]{larochelle1998non}%
  \BibitemOpen
  \bibfield  {author} {\bibinfo {author} {\bibfnamefont {S.}~\bibnamefont
  {Larochelle}}, \bibinfo {author} {\bibfnamefont {A.}~\bibnamefont
  {Talebpour}}, \ and\ \bibinfo {author} {\bibfnamefont {S.-L.}\ \bibnamefont
  {Chin}},\ }\href@noop {} {\bibfield  {journal} {\bibinfo  {journal} {J. Phys.
  B-At. Mol. Opt.}\ }\textbf {\bibinfo {volume} {31}},\ \bibinfo {pages} {1201}
  (\bibinfo {year} {1998})}\BibitemShut {NoStop}%
\bibitem [{\citenamefont {Dundas}\ \emph {et~al.}(1999)\citenamefont {Dundas},
  \citenamefont {Taylor}, \citenamefont {Parker},\ and\ \citenamefont
  {Smyth}}]{dundas99}%
  \BibitemOpen
  \bibfield  {author} {\bibinfo {author} {\bibfnamefont {D.}~\bibnamefont
  {Dundas}}, \bibinfo {author} {\bibfnamefont {K.~T.}\ \bibnamefont {Taylor}},
  \bibinfo {author} {\bibfnamefont {J.~S.}\ \bibnamefont {Parker}}, \ and\
  \bibinfo {author} {\bibfnamefont {E.~S.}\ \bibnamefont {Smyth}},\ }\href@noop
  {} {\bibfield  {journal} {\bibinfo  {journal} {J. Phys. B-At. Mol. Opt.}\
  }\textbf {\bibinfo {volume} {32}},\ \bibinfo {pages} {L231} (\bibinfo {year}
  {1999})}\BibitemShut {NoStop}%
\bibitem [{\citenamefont {Chacon~Salazar}(2014)}]{alexisthesis}%
  \BibitemOpen
  \bibfield  {author} {\bibinfo {author} {\bibfnamefont {A.~A.}\ \bibnamefont
  {Chacon~Salazar}},\ }\emph {\bibinfo {title} {Interaction of attosecond and
  femtosecond pulses with atoms and molecules}},\ \href@noop {} {Ph.D.
  thesis},\ \bibinfo  {school} {Universidad de Salamanca} (\bibinfo {year}
  {2014})\BibitemShut {NoStop}%
\bibitem [{\citenamefont {Chen}\ \emph {et~al.}(2010)\citenamefont {Chen},
  \citenamefont {Ruiz},\ and\ \citenamefont {Becker}}]{chen2010double}%
  \BibitemOpen
  \bibfield  {author} {\bibinfo {author} {\bibfnamefont {S.}~\bibnamefont
  {Chen}}, \bibinfo {author} {\bibfnamefont {C.}~\bibnamefont {Ruiz}}, \ and\
  \bibinfo {author} {\bibfnamefont {A.}~\bibnamefont {Becker}},\ }\href@noop {}
  {\bibfield  {journal} {\bibinfo  {journal} {Phys. Rev. A}\ }\textbf {\bibinfo
  {volume} {82}},\ \bibinfo {pages} {033426} (\bibinfo {year}
  {2010})}\BibitemShut {NoStop}%
\bibitem [{\citenamefont {Hairer}\ \emph {et~al.}(2006)\citenamefont {Hairer},
  \citenamefont {Lubich},\ and\ \citenamefont {Wanner}}]{hairer2006geometric}%
  \BibitemOpen
  \bibfield  {author} {\bibinfo {author} {\bibfnamefont {E.}~\bibnamefont
  {Hairer}}, \bibinfo {author} {\bibfnamefont {C.}~\bibnamefont {Lubich}}, \
  and\ \bibinfo {author} {\bibfnamefont {G.}~\bibnamefont {Wanner}},\
  }\href@noop {} {\emph {\bibinfo {title} {Geometric numerical integration:
  structure-preserving algorithms for ordinary differential equations}}},\
  Vol.~\bibinfo {volume} {31}\ (\bibinfo  {publisher} {Springer-Verlag Berlin
  Heidelberg},\ \bibinfo {year} {2006})\BibitemShut {NoStop}%
\bibitem [{\citenamefont {Efimov}\ \emph {et~al.}(2014)\citenamefont {Efimov},
  \citenamefont {Bezuglov}, \citenamefont {Klyucharev}, \citenamefont {Gnedin},
  \citenamefont {Miculis},\ and\ \citenamefont {Ekers}}]{efimov2014analysis}%
  \BibitemOpen
  \bibfield  {author} {\bibinfo {author} {\bibfnamefont {D.}~\bibnamefont
  {Efimov}}, \bibinfo {author} {\bibfnamefont {N.}~\bibnamefont {Bezuglov}},
  \bibinfo {author} {\bibfnamefont {A.}~\bibnamefont {Klyucharev}}, \bibinfo
  {author} {\bibfnamefont {Y.~N.}\ \bibnamefont {Gnedin}}, \bibinfo {author}
  {\bibfnamefont {K.}~\bibnamefont {Miculis}}, \ and\ \bibinfo {author}
  {\bibfnamefont {A.}~\bibnamefont {Ekers}},\ }\href@noop {} {\bibfield
  {journal} {\bibinfo  {journal} {Opt. Spectrosc.}\ }\textbf {\bibinfo {volume}
  {117}},\ \bibinfo {pages} {8} (\bibinfo {year} {2014})}\BibitemShut {NoStop}%
\bibitem [{\citenamefont {Blanes}\ and\ \citenamefont
  {Moan}(2002)}]{blanesJCAM02}%
  \BibitemOpen
  \bibfield  {author} {\bibinfo {author} {\bibfnamefont {S.}~\bibnamefont
  {Blanes}}\ and\ \bibinfo {author} {\bibfnamefont {P.}~\bibnamefont {Moan}},\
  }\href@noop {} {\bibfield  {journal} {\bibinfo  {journal} {Journal of
  Computational and Applied Mathematics}\ }\textbf {\bibinfo {volume} {142}},\
  \bibinfo {pages} {313} (\bibinfo {year} {2002})}\BibitemShut {NoStop}%
\bibitem [{\citenamefont {Panfili}\ \emph
  {et~al.}(2001{\natexlab{b}})\citenamefont {Panfili}, \citenamefont {Eberly},\
  and\ \citenamefont {Haan}}]{panfiliOE01}%
  \BibitemOpen
  \bibfield  {author} {\bibinfo {author} {\bibfnamefont {R.}~\bibnamefont
  {Panfili}}, \bibinfo {author} {\bibfnamefont {J.~H.}\ \bibnamefont {Eberly}},
  \ and\ \bibinfo {author} {\bibfnamefont {S.~L.}\ \bibnamefont {Haan}},\
  }\href@noop {} {\bibfield  {journal} {\bibinfo  {journal} {Opt. Express}\
  }\textbf {\bibinfo {volume} {8}},\ \bibinfo {pages} {431} (\bibinfo {year}
  {2001}{\natexlab{b}})}\BibitemShut {NoStop}%
\bibitem [{\citenamefont {Panfili}\ \emph
  {et~al.}(2002{\natexlab{b}})\citenamefont {Panfili}, \citenamefont {Haan},\
  and\ \citenamefont {Eberly}}]{panfiliPRL02}%
  \BibitemOpen
  \bibfield  {author} {\bibinfo {author} {\bibfnamefont {R.}~\bibnamefont
  {Panfili}}, \bibinfo {author} {\bibfnamefont {S.~L.}\ \bibnamefont {Haan}}, \
  and\ \bibinfo {author} {\bibfnamefont {J.~H.}\ \bibnamefont {Eberly}},\
  }\href@noop {} {\bibfield  {journal} {\bibinfo  {journal} {Phys. Rev. Lett.}\
  }\textbf {\bibinfo {volume} {89}},\ \bibinfo {pages} {113001} (\bibinfo
  {year} {2002}{\natexlab{b}})}\BibitemShut {NoStop}%
\bibitem [{\citenamefont {Mauger}\ \emph
  {et~al.}(2009{\natexlab{b}})\citenamefont {Mauger}, \citenamefont {Chandre},\
  and\ \citenamefont {Uzer}}]{maugerJPB09}%
  \BibitemOpen
  \bibfield  {author} {\bibinfo {author} {\bibfnamefont {F.}~\bibnamefont
  {Mauger}}, \bibinfo {author} {\bibfnamefont {C.}~\bibnamefont {Chandre}}, \
  and\ \bibinfo {author} {\bibfnamefont {T.}~\bibnamefont {Uzer}},\ }\href@noop
  {} {\bibfield  {journal} {\bibinfo  {journal} {J. Phys. B-At. Mol. Opt.}\
  }\textbf {\bibinfo {volume} {42}},\ \bibinfo {pages} {165602} (\bibinfo
  {year} {2009}{\natexlab{b}})}\BibitemShut {NoStop}%
\bibitem [{\citenamefont {Haan}\ \emph {et~al.}(1994)\citenamefont {Haan},
  \citenamefont {Grobe},\ and\ \citenamefont {Eberly}}]{haan1994numerical}%
  \BibitemOpen
  \bibfield  {author} {\bibinfo {author} {\bibfnamefont {S.~L.}\ \bibnamefont
  {Haan}}, \bibinfo {author} {\bibfnamefont {R.}~\bibnamefont {Grobe}}, \ and\
  \bibinfo {author} {\bibfnamefont {J.~H.}\ \bibnamefont {Eberly}},\ }\href
  {\doibase 10.1103/PhysRevA.50.378} {\bibfield  {journal} {\bibinfo  {journal}
  {Phys. Rev. A}\ }\textbf {\bibinfo {volume} {50}},\ \bibinfo {pages} {378}
  (\bibinfo {year} {1994})}\BibitemShut {NoStop}%
\bibitem [{\citenamefont {Mauger}\ \emph
  {et~al.}(2010{\natexlab{a}})\citenamefont {Mauger}, \citenamefont {Chandre},\
  and\ \citenamefont {Uzer}}]{mauger2010recollisions}%
  \BibitemOpen
  \bibfield  {author} {\bibinfo {author} {\bibfnamefont {F.}~\bibnamefont
  {Mauger}}, \bibinfo {author} {\bibfnamefont {C.}~\bibnamefont {Chandre}}, \
  and\ \bibinfo {author} {\bibfnamefont {T.}~\bibnamefont {Uzer}},\ }\href@noop
  {} {\bibfield  {journal} {\bibinfo  {journal} {Phys. Rev. Lett.}\ }\textbf
  {\bibinfo {volume} {105}},\ \bibinfo {pages} {083002} (\bibinfo {year}
  {2010}{\natexlab{a}})}\BibitemShut {NoStop}%
\bibitem [{\citenamefont {Bhardwaj}\ \emph {et~al.}(2001)\citenamefont
  {Bhardwaj}, \citenamefont {Aseyev}, \citenamefont {Mehendale}, \citenamefont
  {Yudin}, \citenamefont {Villeneuve}, \citenamefont {Rayner}, \citenamefont
  {Ivanov},\ and\ \citenamefont {Corkum}}]{Bhardwaj01}%
  \BibitemOpen
  \bibfield  {author} {\bibinfo {author} {\bibfnamefont {V.~R.}\ \bibnamefont
  {Bhardwaj}}, \bibinfo {author} {\bibfnamefont {S.~A.}\ \bibnamefont
  {Aseyev}}, \bibinfo {author} {\bibfnamefont {M.}~\bibnamefont {Mehendale}},
  \bibinfo {author} {\bibfnamefont {G.~L.}\ \bibnamefont {Yudin}}, \bibinfo
  {author} {\bibfnamefont {D.~M.}\ \bibnamefont {Villeneuve}}, \bibinfo
  {author} {\bibfnamefont {D.~M.}\ \bibnamefont {Rayner}}, \bibinfo {author}
  {\bibfnamefont {M.~Y.}\ \bibnamefont {Ivanov}}, \ and\ \bibinfo {author}
  {\bibfnamefont {P.~B.}\ \bibnamefont {Corkum}},\ }\href {\doibase
  10.1103/PhysRevLett.86.3522} {\bibfield  {journal} {\bibinfo  {journal}
  {Phys. Rev. Lett.}\ }\textbf {\bibinfo {volume} {86}},\ \bibinfo {pages}
  {3522} (\bibinfo {year} {2001})}\BibitemShut {NoStop}%
\bibitem [{\citenamefont {Corkum}(1993)}]{Corkum93}%
  \BibitemOpen
  \bibfield  {author} {\bibinfo {author} {\bibfnamefont {P.~B.}\ \bibnamefont
  {Corkum}},\ }\href {\doibase 10.1103/PhysRevLett.71.1994} {\bibfield
  {journal} {\bibinfo  {journal} {Phys. Rev. Lett.}\ }\textbf {\bibinfo
  {volume} {71}},\ \bibinfo {pages} {1994} (\bibinfo {year}
  {1993})}\BibitemShut {NoStop}%
\bibitem [{\citenamefont {Yudin}\ and\ \citenamefont {Ivanov}(2001)}]{Yudin01}%
  \BibitemOpen
  \bibfield  {author} {\bibinfo {author} {\bibfnamefont {G.~L.}\ \bibnamefont
  {Yudin}}\ and\ \bibinfo {author} {\bibfnamefont {M.~Y.}\ \bibnamefont
  {Ivanov}},\ }\href {\doibase 10.1103/PhysRevA.63.033404} {\bibfield
  {journal} {\bibinfo  {journal} {Phys. Rev. A}\ }\textbf {\bibinfo {volume}
  {63}},\ \bibinfo {pages} {033404} (\bibinfo {year} {2001})}\BibitemShut
  {NoStop}%
\bibitem [{\citenamefont {Chen}\ \emph
  {et~al.}(2018{\natexlab{b}})\citenamefont {Chen}, \citenamefont {Li},
  \citenamefont {Zatsarinny}, \citenamefont {Bartschat},\ and\ \citenamefont
  {Lin}}]{chen2018ratios}%
  \BibitemOpen
  \bibfield  {author} {\bibinfo {author} {\bibfnamefont {Z.}~\bibnamefont
  {Chen}}, \bibinfo {author} {\bibfnamefont {X.}~\bibnamefont {Li}}, \bibinfo
  {author} {\bibfnamefont {O.}~\bibnamefont {Zatsarinny}}, \bibinfo {author}
  {\bibfnamefont {K.}~\bibnamefont {Bartschat}}, \ and\ \bibinfo {author}
  {\bibfnamefont {C.~D.}\ \bibnamefont {Lin}},\ }\href {\doibase
  10.1103/PhysRevA.97.013425} {\bibfield  {journal} {\bibinfo  {journal} {Phys.
  Rev. A}\ }\textbf {\bibinfo {volume} {97}},\ \bibinfo {pages} {013425}
  (\bibinfo {year} {2018}{\natexlab{b}})}\BibitemShut {NoStop}%
\bibitem [{\citenamefont {Ho}\ and\ \citenamefont
  {Eberly}(2005)}]{ho2005classical}%
  \BibitemOpen
  \bibfield  {author} {\bibinfo {author} {\bibfnamefont {P.~J.}\ \bibnamefont
  {Ho}}\ and\ \bibinfo {author} {\bibfnamefont {J.~H.}\ \bibnamefont
  {Eberly}},\ }\href {\doibase 10.1103/PhysRevLett.95.193002} {\bibfield
  {journal} {\bibinfo  {journal} {Phys. Rev. Lett.}\ }\textbf {\bibinfo
  {volume} {95}},\ \bibinfo {pages} {193002} (\bibinfo {year}
  {2005})}\BibitemShut {NoStop}%
\bibitem [{\citenamefont {Ho}\ and\ \citenamefont
  {Eberly}(2006)}]{ho2006plane}%
  \BibitemOpen
  \bibfield  {author} {\bibinfo {author} {\bibfnamefont {P.~J.}\ \bibnamefont
  {Ho}}\ and\ \bibinfo {author} {\bibfnamefont {J.~H.}\ \bibnamefont
  {Eberly}},\ }\href {\doibase 10.1103/PhysRevLett.97.083001} {\bibfield
  {journal} {\bibinfo  {journal} {Phys. Rev. Lett.}\ }\textbf {\bibinfo
  {volume} {97}},\ \bibinfo {pages} {083001} (\bibinfo {year}
  {2006})}\BibitemShut {NoStop}%
\bibitem [{\citenamefont {Mauger}\ \emph
  {et~al.}(2010{\natexlab{b}})\citenamefont {Mauger}, \citenamefont {Chandre},\
  and\ \citenamefont {Uzer}}]{mauger2010fromrecollisions}%
  \BibitemOpen
  \bibfield  {author} {\bibinfo {author} {\bibfnamefont {F.}~\bibnamefont
  {Mauger}}, \bibinfo {author} {\bibfnamefont {C.}~\bibnamefont {Chandre}}, \
  and\ \bibinfo {author} {\bibfnamefont {T.}~\bibnamefont {Uzer}},\ }\href@noop
  {} {\bibfield  {journal} {\bibinfo  {journal} {Phys. Rev. Lett.}\ }\textbf
  {\bibinfo {volume} {104}},\ \bibinfo {pages} {043005} (\bibinfo {year}
  {2010}{\natexlab{b}})}\BibitemShut {NoStop}%
\bibitem [{\citenamefont {Brics}\ \emph {et~al.}(2014)\citenamefont {Brics},
  \citenamefont {Rapp},\ and\ \citenamefont {Bauer}}]{brics2014nonsequential}%
  \BibitemOpen
  \bibfield  {author} {\bibinfo {author} {\bibfnamefont {M.}~\bibnamefont
  {Brics}}, \bibinfo {author} {\bibfnamefont {J.}~\bibnamefont {Rapp}}, \ and\
  \bibinfo {author} {\bibfnamefont {D.}~\bibnamefont {Bauer}},\ }\href@noop {}
  {\bibfield  {journal} {\bibinfo  {journal} {Phys. Rev. A}\ }\textbf {\bibinfo
  {volume} {90}},\ \bibinfo {pages} {053418} (\bibinfo {year}
  {2014})}\BibitemShut {NoStop}%
\bibitem [{\citenamefont {Parker}\ \emph {et~al.}(2007)\citenamefont {Parker},
  \citenamefont {Meharg}, \citenamefont {McKenna},\ and\ \citenamefont
  {Taylor}}]{parker2007single}%
  \BibitemOpen
  \bibfield  {author} {\bibinfo {author} {\bibfnamefont {J.~S.}\ \bibnamefont
  {Parker}}, \bibinfo {author} {\bibfnamefont {K.~J.}\ \bibnamefont {Meharg}},
  \bibinfo {author} {\bibfnamefont {G.~A.}\ \bibnamefont {McKenna}}, \ and\
  \bibinfo {author} {\bibfnamefont {K.~T.}\ \bibnamefont {Taylor}},\
  }\href@noop {} {\bibfield  {journal} {\bibinfo  {journal} {J. Phys. B-At.
  Mol. Opt.}\ }\textbf {\bibinfo {volume} {40}},\ \bibinfo {pages} {1729}
  (\bibinfo {year} {2007})}\BibitemShut {NoStop}%
\bibitem [{\citenamefont {Ilkov}\ \emph {et~al.}(1992)\citenamefont {Ilkov},
  \citenamefont {Decker},\ and\ \citenamefont {Chin}}]{ilkov1992ionization}%
  \BibitemOpen
  \bibfield  {author} {\bibinfo {author} {\bibfnamefont {F.}~\bibnamefont
  {Ilkov}}, \bibinfo {author} {\bibfnamefont {J.}~\bibnamefont {Decker}}, \
  and\ \bibinfo {author} {\bibfnamefont {S.}~\bibnamefont {Chin}},\ }\href@noop
  {} {\bibfield  {journal} {\bibinfo  {journal} {J. Phys. B-At. Mol. Opt.}\
  }\textbf {\bibinfo {volume} {25}},\ \bibinfo {pages} {4005} (\bibinfo {year}
  {1992})}\BibitemShut {NoStop}%
\bibitem [{\citenamefont {l'Huillier}\ \emph {et~al.}(1983)\citenamefont
  {l'Huillier}, \citenamefont {Lompre}, \citenamefont {Mainfray},\ and\
  \citenamefont {Manus}}]{huilier1983multiply}%
  \BibitemOpen
  \bibfield  {author} {\bibinfo {author} {\bibfnamefont {A.}~\bibnamefont
  {l'Huillier}}, \bibinfo {author} {\bibfnamefont {L.~A.}\ \bibnamefont
  {Lompre}}, \bibinfo {author} {\bibfnamefont {G.}~\bibnamefont {Mainfray}}, \
  and\ \bibinfo {author} {\bibfnamefont {C.}~\bibnamefont {Manus}},\ }\href
  {\doibase 10.1103/PhysRevA.27.2503} {\bibfield  {journal} {\bibinfo
  {journal} {Phys. Rev. A}\ }\textbf {\bibinfo {volume} {27}},\ \bibinfo
  {pages} {2503} (\bibinfo {year} {1983})}\BibitemShut {NoStop}%
\bibitem [{\citenamefont {Huismans}\ \emph {et~al.}(2011)\citenamefont
  {Huismans}, \citenamefont {Rouz{\'e}e}, \citenamefont {Gijsbertsen},
  \citenamefont {Jungmann}, \citenamefont {Smolkowska}, \citenamefont {Logman},
  \citenamefont {Lepine}, \citenamefont {Cauchy}, \citenamefont {Zamith},
  \citenamefont {Marchenko} \emph {et~al.}}]{huismans2011time}%
  \BibitemOpen
  \bibfield  {author} {\bibinfo {author} {\bibfnamefont {Y.}~\bibnamefont
  {Huismans}}, \bibinfo {author} {\bibfnamefont {A.}~\bibnamefont
  {Rouz{\'e}e}}, \bibinfo {author} {\bibfnamefont {A.}~\bibnamefont
  {Gijsbertsen}}, \bibinfo {author} {\bibfnamefont {J.}~\bibnamefont
  {Jungmann}}, \bibinfo {author} {\bibfnamefont {A.}~\bibnamefont
  {Smolkowska}}, \bibinfo {author} {\bibfnamefont {P.}~\bibnamefont {Logman}},
  \bibinfo {author} {\bibfnamefont {F.}~\bibnamefont {Lepine}}, \bibinfo
  {author} {\bibfnamefont {C.}~\bibnamefont {Cauchy}}, \bibinfo {author}
  {\bibfnamefont {S.}~\bibnamefont {Zamith}}, \bibinfo {author} {\bibfnamefont
  {T.}~\bibnamefont {Marchenko}},  \emph {et~al.},\ }\href@noop {} {\bibfield
  {journal} {\bibinfo  {journal} {Science}\ }\textbf {\bibinfo {volume}
  {331}},\ \bibinfo {pages} {61} (\bibinfo {year} {2011})}\BibitemShut
  {NoStop}%
\bibitem [{\citenamefont {Shaaran}\ \emph {et~al.}(2012)\citenamefont
  {Shaaran}, \citenamefont {Figueira~de Morisson~Faria},\ and\ \citenamefont
  {Schomerus}}]{shaaran2012causality}%
  \BibitemOpen
  \bibfield  {author} {\bibinfo {author} {\bibfnamefont {T.}~\bibnamefont
  {Shaaran}}, \bibinfo {author} {\bibfnamefont {C.}~\bibnamefont {Figueira~de
  Morisson~Faria}}, \ and\ \bibinfo {author} {\bibfnamefont {H.}~\bibnamefont
  {Schomerus}},\ }\href {\doibase 10.1103/PhysRevA.85.023423} {\bibfield
  {journal} {\bibinfo  {journal} {Phys. Rev. A}\ }\textbf {\bibinfo {volume}
  {85}},\ \bibinfo {pages} {023423} (\bibinfo {year} {2012})}\BibitemShut
  {NoStop}%
\bibitem [{\citenamefont {Ayadi}\ \emph {et~al.}(2017)\citenamefont {Ayadi},
  \citenamefont {F\"oldi}, \citenamefont {Dombi},\ and\ \citenamefont
  {T\"ok\'esi}}]{Ayadi17}%
  \BibitemOpen
  \bibfield  {author} {\bibinfo {author} {\bibfnamefont {V.}~\bibnamefont
  {Ayadi}}, \bibinfo {author} {\bibfnamefont {P.}~\bibnamefont {F\"oldi}},
  \bibinfo {author} {\bibfnamefont {P.}~\bibnamefont {Dombi}}, \ and\ \bibinfo
  {author} {\bibfnamefont {K.}~\bibnamefont {T\"ok\'esi}},\ }\href
  {http://stacks.iop.org/0953-4075/50/i=8/a=085005} {\bibfield  {journal}
  {\bibinfo  {journal} {J. Phys. B-At. Mol. Opt.}\ }\textbf {\bibinfo {volume}
  {50}},\ \bibinfo {pages} {085005} (\bibinfo {year} {2017})}\BibitemShut
  {NoStop}%
\end{thebibliography}%

\end{document}